\documentclass[12pt,a4paper]{article}
\usepackage[T1]{fontenc}
\usepackage[utf8]{inputenc}
\usepackage{authblk}
\usepackage{bbding} 

\usepackage{amsmath}
\usepackage{graphicx,float,color}
\usepackage{hyperref}
\hypersetup{colorlinks = true, linkcolor = blue, urlcolor  = blue, citecolor = blue, anchorcolor = blue}
\usepackage{multirow}
\usepackage[numbers,sort&compress]{natbib}
\usepackage{braket}
\usepackage{amsfonts}
\usepackage{amssymb} 
\usepackage{epic}
\usepackage{eepic}
\usepackage{epsfig}
\usepackage{latexsym}
\usepackage{color}
\usepackage{url}  
\usepackage{caption}
\usepackage[caption=false]{subfig}
\usepackage{float}

\usepackage{bm}
\usepackage{geometry}
\usepackage{slashed}
\usepackage[detect-all]{siunitx}

\usepackage{authblk}
\usepackage{physics}
\usepackage{amsmath}

\usepackage{amsfonts}
\usepackage{color}
\usepackage{url}
\usepackage{graphicx,float,color}
\usepackage{verbatim}

\usepackage{tablefootnote}
\usepackage{lipsum}
\makeatletter
\def\ps@pprintTitle{%

 \let\@oddhead\@empty
 \let\@evenhead\@empty
 \def\@oddfoot{}%
 \let\@evenfoot\@oddfoot}
\makeatother
\usepackage{etoolbox}
\def\bh{\mathrm{BH}}
\def\uco{\mathrm{UCO}}
\def\echo{\mathrm{echo}}
\def\out{\mathrm{out}}
\def\in{\mathrm{in}}
\def\tra{\mathrm{trans}}
\def\re{\mathrm{ref}}
\def\wa{\mathrm{wall}}
\def\lr{\mathrm{LR}}
\def\bx{\bar{x}}
\def\bA{\bar{A}}
\def\dir{\mathrm{dir}}
\def\qnm{\mathrm{QNM}}
\def\obs{\mathrm{obs}}

\begin{document}
\title{\textbf{Causal Green function decomposition for quantum black hole seismology}}

\author[1,2]{Xi-Li Zhang\thanks{zhangxili@ihep.ac.cn}}
\author[1,3]{Jing Ren\thanks{renjing@ihep.ac.cn}}

\affil[1]{\normalsize Institute of High Energy Physics, Chinese Academy of Sciences, Beijing 100049, China}
\affil[2]{\normalsize School of Physics Sciences, University of Chinese Academy of Sciences, Beijing 100039, China}
\affil[3]{\normalsize Center for High Energy Physics, Peking University, Beijing 100871, China}
\maketitle

%\abstract{ 

\begin{abstract}

The growing sensitivity of gravitational-wave detectors enables increasingly precise tests of black hole (BH) ringdown spectroscopy. BH quasinormal modes (QNMs) are, however, spectrally unstable: small near-horizon modifications can produce a drastically different QNM spectrum, while causality requires the prompt ringdown to remain BH-like until the reflected signal returns. Quantum BHs with substantial interior reflection provide a natural setting for this tension, yet the relation between their time-domain waveform and different QNM spectra still lacks a consistent picture. 
In this work, we systematically examine the time-domain Green function for quantum BHs, considering sources located outside and inside the light-ring potential barrier. By decomposing the Green function into causally distinct components and choosing the corresponding inverse-Laplace contours consistently, we clarify how the response is built from different sets of QNMs. We find that the quantum BH QNM reconstruction always faithfully describes the signal once the curved spacetime is probed, but its practical efficiency depends strongly on the evolutionary stage. Before interior reflection becomes relevant, we prove that this basis is formally equivalent to the BH QNM and tail expansions, with convergence properties sensitive to source location. At late times, the long-lived modes provide an efficient basis. Time-domain simulations confirm these results, providing a unified causal framework for BH spectroscopy and quantum BH seismology.

\end{abstract}

\newpage
{
  \hypersetup{linkcolor=black}
  \tableofcontents
}

%\begin{document}
%\maketitle
%\flushbottom

\section{Introduction}
\label{sec:intro}

The first gravitational wave detection of a compact binary coalescence, GW150914, by LIGO in 2015~\cite{LIGOScientific:2016aoc}, opened the era of gravitational wave astronomy. Since then, the ground based LIGO-Virgo-KAGRA (LVK) network has made remarkable progress, recently reporting more than 300 confirmed events in the GWTC-5.0 catalogue~\cite{LIGOScientific:2026sit,LIGOScientific:2026wfs}. In particular, the loudest event, GW250114, is remarkably similar to GW150914, but has a network signal-to-noise ratio (SNR) of nearly 80, more than three times that of GW150914, thanks to the improved O4 sensitivity. This event enables precise black hole (BH) spectroscopy: two distinct quasinormal modes (QNMs) are independently identified in the ringdown stage. The measured frequencies and damping times are consistent with the Kerr BH prediction for the $\ell=|m|=2$ fundamental mode and its first overtone, providing a test of the Kerr nature of the merger remnant~\cite{LIGOScientific:2025rid,LIGOScientific:2025wao}.

However, to what extent does this test verify the BH nature of the remnant? If a future, even louder event enables precise measurements of many BH QNMs and the late-time tail~\cite{Price:1971fb,Price:1972pw}, would this confirm the defining feature of a BH, i.e. the event horizon? In BH perturbation theory, BH QNMs are known to suffer from spectral instability (see the review~\cite{Berti:2025hly}): a tiny change of the potential or boundary condition can drastically alter the spectrum. Causality, however, dictates that if such a change is confined sufficiently far from the light-ring potential barrier, its effect can reach a distant observer only after a finite delay. The early prompt ringdown should therefore remain indistinguishable from the BH prediction. These two facts stand in tension.
%These two perspectives give rise to what may be called a causality dilemma.    

A well motivated physical realization of this tension is the near-horizon corrections expected from quantum gravity, which may lead to quantum BHs or  horizonless ultracompact objects (UCOs)~\cite{Mathur:2005zp,Holdom:2016nek,Holdom:2019ouz,Ren:2019afg,Oshita:2019sat}. Such corrections could potentially resolve various theoretical problems associated with BHs, while also modifying the ingoing boundary condition at the horizon (see the review \cite{Cardoso:2019rvt}). For a distant observer, substantial interior reflection would produce gravitational wave echoes in the post-merger stage, offering a smoking-gun signature of quantum gravity effects around astrophysical BHs~\cite{Cardoso:2016rao,Cardoso:2016oxy}. The resulting quantum BH or UCO has a QNM spectrum very different from that of a classical BH, including a distinct set of long-lived modes close to the real axis. This motivates ``quantum BH seismology'' in the post-merger stage~\cite{Conklin:2017lwb,Oshita:2020dox}--extracting the quantum BH spectrum to infer its properties, rather than ordinary ``BH spectroscopy''~\cite{Dreyer:2003bv,Berti:2005ys}.  Time domain simulations, however, reveal that the prompt ringdown remains essentially identical to the classical BH prediction for times shorter than the interior reflection delay $t_d$. How can the early signal be unchanged when the underlying QNM spectrum is so drastically different? What, then, does ``quantum BH seismology'' actually mean?

A resolution to this puzzle appears well known: the Green function for a quantum BH or UCO can be expanded as the standard BH Green function plus an echo correction~\cite{Mark:2017dnq,Nakano:2017fvh,Correia:2018apm,Wang:2019rcf,Hui:2019aox}. The echo part can be further expanded as a geometric series, each term corresponding to a successive pulse with increasing delay by $t_d$. It is then tempting to identify a finite time observation with a finite truncation of that series. In such an expansion, however, the original quantum BH QNMs disappear entirely, making the naive notion of ``quantum BH seismology'' seemingly irrelevant~\cite{Daghigh:2025wcw} -- suggesting that the apparent resolution is itself incomplete. %Although the relation between the time-domain waveform and different sets of QNMs has been examined in numerous studies, a rigorous and consistent picture of how these approaches fit together is still missing. 
On the other hand, even for standard BH spectroscopy, whose foundation was laid by Leaver's seminal work~\cite{Leaver:1986gd}, it has only recently become clear when the BH QNM expansion faithfully describes the time-domain Green function~\cite{Chavda:2024awq,DeAmicis:2025xuh,Arnaudo:2025uos,Arnaudo:2025kit,Su:2026fvj,Rosato:2026moe}. The key insight is to properly decompose the Green function at each stage of the evolution and to evaluate the individual components using appropriately chosen contour integrals that respect causality~\cite{Arnaudo:2025uos,Arnaudo:2025kit,Su:2026fvj}.

In this work, we examine the time-domain Green function for quantum BHs in order to provide a consistent picture of how its temporal evolution is related to different sets of QNMs. Although this relation has been explored from many angles, how these approaches fit together remains unclear. 
Specifically, we extend the recently developed BH spectroscopy methodology to the quantum BH case, considering two representative configurations where the source is placed either outside or inside the light ring. Owing to interior reflection, the causal decomposition of the Green function for quantum BHs exhibits novel features compared to the classical BH case. This framework then provides a solid foundation not only for a physical interpretation of the evolution at different time scales, with the causality requirement explicitly incorporated, but also for a clear resolution of the tension between early-time BH ringdown and quantum BH seismology. We also perform numerical simulations to verify the theoretical framework.

This paper is structured as follows. In Sec.~\ref{sec:theory} we develop the theoretical framework for the causal decomposition of the time-domain Green function in the two cases where the source is located outside and inside the light ring. Section~\ref{sec:validaiton} presents numerical simulations that verify this framework, demonstrating agreement between the time-domain evolution and the various QNM reconstructions. We discuss the implications and summarize our conclusions in Sec.~\ref{sec:summary}. Additional details on the asymptotic and analytic properties of the homogeneous solutions, and on the impact of initial conditions on the QNM reconstruction, are provided in Appendices~\ref{app:homosolution} and~\ref{app:initialcond}, respectively. 
Throughout the paper we adopt geometric units, $G=c=1$, unless stated otherwise.

\section{Theoretical framework}
\label{sec:theory}

To examine the properties of the Green function for quantum BHs or horizonless UCOs, we focus on the non-rotating case in this work. We adopt a simple phenomenological model for these objects: a truncated BH with a reflective surface placed just outside the would-be horizon~\cite{Conklin:2017lwb,Mark:2017dnq,Wang:2019rcf}. Although simplified, this model captures a wide range of near-horizon corrections through the choice of the surface location and its effective reflectivity, offering an efficient phenomenological framework for the Green function study. For the convenience of later discussion, we will not distinguish between quantum BHs and UCOs. Instead, we use ``UCO'' to denote all candidates with substantial interior reflection.

For the simplified model, the spacetime exterior to the surface radius $r_0$ is described by the Schwarzschild metric, and linear perturbations are governed by the one-dimensional radial wave equation:
\begin{equation}\label{eq:EOMtime}
    \left( \frac{\partial^2}{\partial x^2} -\frac{\partial^2}{\partial t^2} - V(r) \right) \psi(t, x) = 0,
\end{equation}
where $\psi$ is the master variable and $x$ is the tortoise coordinate implicitly defined by $dx/dr=(1-2M/r)^{-1}$. 
$V(r)$ is the effective potential. Although it takes a slightly different form for different spin or perturbation sectors, it is generally characterized by a potential barrier with a peak around the light ring radius $r_\lr=3M$ (with the tortoise coordinate denoted by $x_\lr$), and it approaches zero both at spatial infinity ($x\to\infty$) and at the horizon ($x\to -\infty$). For UCOs, the surface location $x_0$ is chosen in the negative asymptotic regime, so that the only difference from the BH case is the nonzero reflection at the surface, parameterized by a function $R_\wa(\omega)$. For later discussion, it is convenient to define  a time scale associated with the position of the reflection surface, namely, the time delay $t_d\equiv 2(x_\lr-x_0)$ for the round trip time between the potential barrier and the inner surface. 

Considering an initial-value problem, the time-domain waveform $\psi(t,x)$ can be obtained from the retarded time-domain Green function $G(t,x,x')$ via
\begin{eqnarray}\label{eq:timedomwf}
    \psi(t,x)=-\int^\infty_{-\infty} \left[\partial_t G(t,x,x')\psi(t_0,x')+G(t,x,x')\partial_t \psi(t_0,x')\right]dx'\,,
\end{eqnarray}
which relates the late-time waveform to its initial conditions at $t=t_0$. The retarded Green function satisfies $G(t,x,x')=0$ for $t<t_0$, while for $t\geq t_0$ it can be obtained from the frequency-domain Green function via the inverse Laplace transform,
\begin{eqnarray}\label{eq:GtInvLap}
  G(t,x,x')=\frac{1}{2\pi}\int_{-\infty+i\epsilon}^{\infty+i\epsilon} \tilde{G}(\omega,x,x')e^{-i\omega t} d\omega\,,
\end{eqnarray}
where $\epsilon>0$ ensures the correct causal contour. For an asymptotic observer at large distance, we can assume a compact source with $x'< x$ without loss of generality. The frequency-domain Green function then reads
\begin{eqnarray}\label{eq:GFomega0}
   \tilde{G}(\omega, x,x')=\frac{1}{W(\omega)}X^+(\omega, x)X^-(\omega, x')\,.
\end{eqnarray}
Here, $X^+(\omega, x)$ and $X^-(\omega, x)$ denote the two linearly independent solutions satisfying the required boundary conditions at spatial infinity and near the horizon, respectively. The Wronskian $W(X^+,X^-)=\partial_x X^+ X^--X^+\partial_x X^-$ is independent of $x$, and we denote it by $W(\omega)$.

The inverse Laplace transform in Eq.~(\ref{eq:GtInvLap}) can be evaluated by deforming the integration contour in the complex-$\omega$ plane. 
For the BH case, Leaver's original contour-integral formulation yields the standard BH spectroscopy program, with distinct contributions from QNMs, branch cuts, and the large arc~\cite{Leaver:1986gd}. This approach, however, leads to issues such as diverging QNM contributions at early times and difficulty in computing the direct wave part. Recently, more careful examination has shown that both issues can be resolved by evaluating the waveform using a different contour that respects causality~\cite{Arnaudo:2025uos,Arnaudo:2025kit,Su:2026fvj}. 
Below, we generalize this new methodology to evaluate the time-domain  Green function $G_\uco(t,x,x')$ for UCOs in two representative source configurations. We first consider a source located far outside the light ring, corresponding to perturbations by small compact objects in the exterior region. We also consider a source well inside the light ring, motivated by potential disturbances originating from the core of a newly forming UCO from compact binary coalescence.

\subsection{Source outside the light ring}
\label{sec:theory_outside}

To set the stage, we first review the new methodology for decomposing the BH Green function, following Ref.~\cite{Su:2026fvj} for a source outside the light ring ($x\gtrsim x'\gg x_\lr$). 
%To identify the appropriate decomposition, it is instructive to examine the behavior of the Green function at 
The appropriate decomposition can be identified from the 
large-$|\omega|$ behavior of the Green function in the complex-$\omega$ plane. In this limit, Eq.~(\ref{eq:GFomega0}) simplifies when the asymptotic expansions of the solutions are substituted. For BHs, inserting Eqs.~(\ref{eq:SNEampp}) and (\ref{eq:SNEampm}) into Eq.~(\ref{eq:GFomega0}) yields
\begin{eqnarray}
    \tilde{G}_\bh(\omega, x,x')&=&\frac{1}{W_\bh(\omega)}X_\bh^+(\omega, x)X_\bh^-(\omega, x')
    %=\frac{\left(A_\in^\bh e^{-i\omega x'}+A_\out^\bh e^{i\omega x'}\right)D_\out^\bh e^{i\omega x}}{A_\in^\bh D_\out^\bh}\nonumber\\
    \to \frac{1}{2i\omega}\left[e^{i\omega (x-x')}+\frac{A_\out^\bh(\omega)}{A_\in^\bh(\omega)}e^{i\omega (x+x')}\right],
\end{eqnarray}
where $A_\in^\bh(\omega)$ and $A_\out^\bh(\omega)$ are asymptotic amplitudes of $X^-_\bh(\omega, x)$ at spatial infinity. The Wronskian is given by $W_\bh(\omega)=2i\omega A_{\in}^\bh(\omega)$, and the zeros of $A_{\in}^\bh(\omega)$ define the BH QNMs.

The two terms in brackets have distinct physical meanings and are associated with two different time scales. The first term corresponds to the direct wave propagating straight from the source to the observer, characterized by the time scale
\begin{eqnarray}\label{eq:tdir}
    t_\dir(x,x')\equiv x-x'\,.
\end{eqnarray}
The second term represents the reflection from the light-ring potential barrier in the strong-gravity regime.  Since the reflection occurs near the peak of the potential barrier at $x_\lr$, it is useful to factor out the $x_\lr$ dependence from the BH amplitudes by writing $A^\bh_\out(\omega)=\bA^\bh_\out(\omega) e^{-2i\omega x_\lr}$. The second term can then be rewritten as $\bA^\bh_\out(\omega)/A^\bh_\in(\omega)e^{i\omega(\bx+\bx')}$, where $\bx^{(\prime)}=x^{(\prime)}-x_\lr$. This then define the second time scale\footnote{The $x_\lr$ dependence of $t_\lr$,  which is implicit in Ref.~\cite{Su:2026fvj}, is discussed explicitly in Refs.~\cite{Chavda:2024awq,Rosato:2026moe}.}  
\begin{eqnarray}\label{eq:tlr}
    t_\lr(x,x')\equiv \bx+\bx'\,.
\end{eqnarray}
Motivated by this asymptotic form, the ingoing BH solution can be decomposed as $X^-_\bh(\omega, x')=A_\in^\bh(\omega) X^{\prime +}_\bh(\omega, x')+A_\out^\bh(\omega) X^+_\bh(\omega, x')$, where $X_\bh^{\prime +}$ denotes the ingoing solution at spatial infinity. The Green function at a generic frequency $\omega$ can then be decomposed into two components~\cite{Su:2026fvj}:
\begin{eqnarray}\label{eq:GFomega1}
    \tilde{G}_\bh(\omega, x,x')=\tilde{G}^-_\bh(\omega, x,x')+\tilde{G}^+_\bh(\omega, x,x')\,,
\end{eqnarray}
where 
\begin{eqnarray}\label{eq:GBHpm}
    \tilde{G}^-_\bh(\omega, x,x')=\frac{X_\bh^+(\omega, x)X_\bh^{\prime +}(\omega, x')}{2i\omega},\quad \tilde{G}^+_\bh(\omega, x,x')=\frac{A_\out^\bh(\omega)}{A_\in^\bh(\omega)}\frac{X_\bh^+(\omega, x)X_\bh^+(\omega, x')}{2i\omega}\,.
\end{eqnarray}

Substituting the decomposition Eq.~(\ref{eq:GFomega1}) into the inverse Laplace transform Eq.~(\ref{eq:GtInvLap}), we see that in the large-$|\omega|$ limit the contributions of $\tilde{G}^-_\bh$ and $\tilde{G}^+_\bh$ carry the phase factors $e^{-i\omega(t-t_\dir(x,x'))}$ and $e^{-i\omega(t-t_\lr(x,x'))}$, respectively. To ensure convergence of the large-arc contribution for each component, the contour must be closed in the upper or lower half-plane depending on how $t$ compares with its associated time scale.  
For $t < t_\dir(x,x')$, both components are closed in the upper half-plane, where their sum is analytic and yields zero. For $t_\dir(x,x') < t < t_\lr(x,x')$, $\tilde{G}^-_\bh$ is closed in the lower half-plane, while $\tilde{G}^+_\bh$ is closed in the upper half-plane. This yields the direct wave contribution, which is dominated by $\tilde{G}^-_\bh$ and  is therefore insensitive to the curved-spacetime information in $\tilde{G}^+_\bh$~\cite{Su:2026fvj}. For $t > t_\lr(x,x')$, the full Green function $\tilde{G}_\bh$ is enclosed in the lower half-plane, as in Leaver's original contour. The poles and branch cuts in this half-plane yield the standard BH QNM and tail contributions.

This decomposition is consistent with the analytical findings for the toy model of a delta-function potential in Ref.~\cite{Chavda:2024awq}. There, the counterpart of $\tilde{G}^+_\bh$ is associated with a step function $\Theta(t-t_\lr(x,x'))$, so that this part contributes only after the characteristic time $t_\lr(x,x')$. In the more realistic case studied here, where the potential decays slowly at large distances, the small contribution from $\tilde{G}^+_\bh$ at $t<t_\lr(x,x')$ arises from the branch cut in the upper half-plane, a direct consequence of that slow decay.

\begin{figure}[htbp] 
%    \centering 
       
    \includegraphics[width=0.75\textwidth]{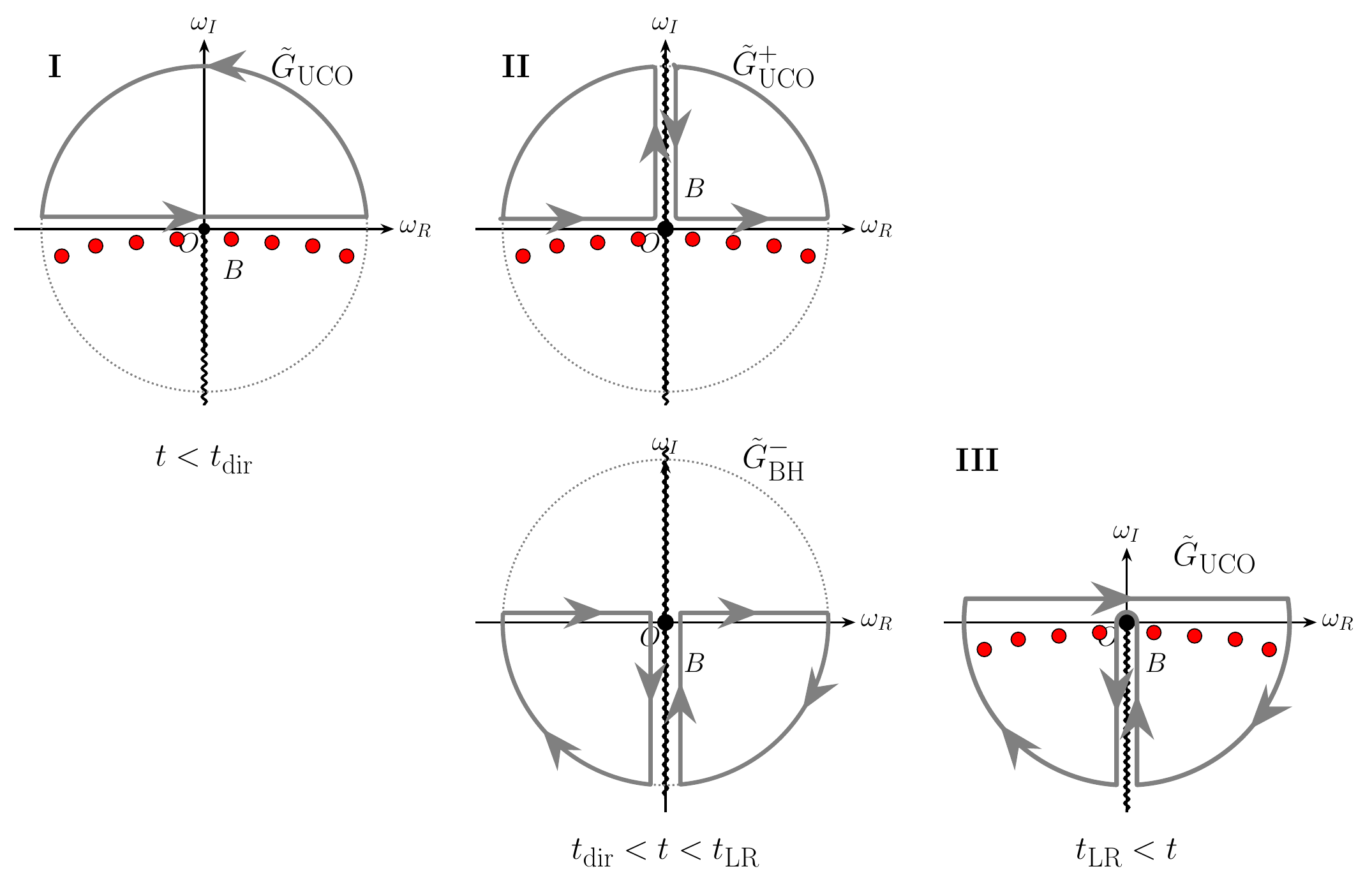}

    \caption{Decomposition of the UCO Green function and appropriate enclosing contours in the complex-$\omega$ plane  for the inverse Laplace transform at different evolution stages, for a source outside the light ring. The three stages are separated by the characteristic time scales $t_\dir(x,x')$ and $t_\lr(x,x')$ (see Eqs.~(\ref{eq:tdir}) and (\ref{eq:tlr})), denoted simply as $t_\dir$ and $t_\lr$ in the plot). Red circles mark the UCO QNMs; $B$ denotes the branch cuts.}
    %Evolution of the integration contours in the complex $\omega$-plane for evaluating the time-domain Green's functions at different evolutionary stages.}
    \label{fig:contour_outside} 
\end{figure}

We now extend the discussion to the UCO Green function. Compared to the BH case, the solution $X^-_\bh(\omega, x')$ is replaced by $X^-_\uco(\omega, x')$. Taking the large-$|\omega|$ limit and substituting the asymptotic expansions from Eqs.~(\ref{eq:SNEampp}) and (\ref{eq:SNEamppUCO}) into Eq.~(\ref{eq:GFomega0}) yields
\begin{eqnarray}\label{eq:GUCOdecLw}
    \tilde{G}_\uco(\omega, x,x')=\frac{1}{W_\uco(\omega)}X_\bh^+(\omega, x)X_\uco^-(\omega, x')
    \to \frac{1}{2i\omega}\left[e^{i\omega (x-x')}+\frac{A_\out^\uco(\omega)}{A_\in^\uco(\omega)}e^{i\omega (x+x')}\right].
\end{eqnarray}
The UCO Wronskian is $W_\uco(\omega)=2i\omega A_{\in}^\uco(\omega)$, and the UCO QNMs $\omega_n$ are determined by $A_\in^\uco(\omega_n)=0$. Using the relation between the UCO and BH amplitudes, this condition translates into (see the derivation of Eq.~(\ref{eq:UCOQNMs_app}) in Appendix~\ref{app:homosolution})
\begin{eqnarray}\label{eq:UCOQNMs}
    1-R_{\bh}(\omega_n)R_\wa(\omega_n) e^{i\omega_n t_d}=0\,,
\end{eqnarray}
where $R_\bh(\omega)$ is the reflection coefficient of the light-ring potential barrier defined in Eq.~(\ref{eq:RBH}) and $R_\wa(\omega)$ denotes the interior reflection given in Eq.~(\ref{eq:Aucoref}).
Separating this equation into real and imaginary parts gives two conditions for UCO QNMs,
\begin{align}\label{eq:UCOQNMs2}
    t_d \omega_{n,R} = 2n\pi + \phi_n,\quad
    t_d \omega_{n,I} = \ln \left| R_\bh(\omega_n) R_\wa(\omega_n) \right|\,,
\end{align}
where $\phi_n = \arg\!\left[ R_\bh(\omega_n) R_\wa(\omega_n) \right]$. Stability requires $\omega_{n,I}<0$, i.e. $|R_{\bh}(\omega)R_\wa(\omega)|<1$. 
When the interior reflection is strong, the effective reflectivity $|R_{\bh}(\omega)R_\wa(\omega)|$ can approach unity at frequencies below the BH fundamental mode $\omega_{\textrm{RD},R}$. This gives rise to a distinct set of long-lived (trapped) modes that differ drastically from the rapidly decaying QNMs of a BH.

As in the BH case, after extracting the $x_\lr$ dependence,  the two terms in Eq.~(\ref{eq:GUCOdecLw}) are again associated with the time scales $t_\dir(x,x')$ and $t_\lr(x,x')$ in Eqs.~(\ref{eq:tdir}) and (\ref{eq:tlr}). The only difference is that the ratio $\bA_\out/A_\in$ for the BH is replaced by the corresponding ratio for the UCO. The Green function then admits a similar decomposition: using $X^-_\uco(\omega,x')=A_\in^\uco(\omega) X^{\prime +}_\bh(\omega,x')+A_\out^\uco(\omega) X^+_\bh(\omega,x')$, we find for a generic frequency
\begin{eqnarray}\label{eq:GFomegaUCO1}
    \tilde{G}_\uco(\omega, x,x')&=&\tilde{G}^-_\bh(\omega, x,x')+\tilde{G}^+_\uco(\omega, x,x')\,,
\end{eqnarray}
with $\tilde{G}^-_\bh(\omega, x,x')$ given in Eq.~(\ref{eq:GBHpm}) and
\begin{eqnarray}\label{eq:tGucopOUT}
    \tilde{G}^+_\uco(\omega, x,x')&=&\frac{A_\out^\uco(\omega)}{A_\in^\uco(\omega)}\frac{X_\bh^+(\omega, x)X_\bh^+(\omega, x')}{2i\omega}\,.
\end{eqnarray}
Following the same reasoning, the UCO Green function can be evaluated at different times as illustrated in Fig.~\ref{fig:contour_outside}, in accordance with causality. 
In Stage I, $\tilde{G}_\uco$ is analytic in the upper half-plane, just as in the BH case, so no signal appears.  
In Stage II, we again close the contour in the lower half-plane for $\tilde{G}^-_\bh$ and in the upper half-plane for $\tilde{G}^+_\uco$. The first term gives the same dominant direct wave contribution as for a BH. The branch cut of the UCO in the upper half-plane turns out to be identical to that of the BH,\footnote{As we will show later, this follows from the fact that $\tilde{G}^+_\echo$ in Eq.~(\ref{eq:Gpucodec}), i.e. the difference between $\tilde{G}^+_\uco$ and $\tilde{G}^+_\bh$, is analytic in the upper half-plane, and thus does not contribute any additional branch cut structure.} and so no UCO corrections appear at this stage. 
In Stage III, the UCO Green function receives contributions from its poles and branch cuts in the lower half-plane, and is therefore governed by the UCO QNMs and tail. This forms the basis of ``quantum BH seismology''~\cite{Conklin:2017lwb,Oshita:2020dox}.

The tension mentioned earlier now becomes manifest. On the one hand, the UCO QNMs given by Eq.~(\ref{eq:UCOQNMs2}) are fundamentally different from their BH counterparts. On the other hand, when $t_d$ is sufficiently large compared to the typical BH QNM ringing timescale $\mathcal{O}(M)$, causality demands that the time-domain waveform closely follows the BH prediction before the characteristic echo time
\begin{eqnarray}\label{eq:techo}
    t_\echo(x,x') \equiv t_\lr(x,x') + t_d\,.
\end{eqnarray}
How can the UCO QNMs and tail contributions then exactly reproduce the BH result? 
A resolution has been proposed in the literature~\cite{Mark:2017dnq,Correia:2018apm,Wang:2019rcf,Hui:2019aox}: the UCO amplitude ratio can be expanded as the BH part plus interior-reflection corrections. Specifically, using Eq.~(\ref{eq:Rel21}) for $\omega$ away from the BH or UCO QNMs, the UCO Green function $\tilde{G}^+_\uco(\omega, x,x')$ can be decomposed as
\begin{eqnarray}\label{eq:Gpucodec}
    \tilde{G}^+_\uco(\omega, x,x')
    =\tilde{G}^+_\bh(\omega, x,x')+\tilde{G}^+_\echo(\omega, x,x')\,,
\end{eqnarray}
where $\tilde{G}^+_\bh(\omega, x,x')$ is given in Eq.~(\ref{eq:GBHpm}) and
\begin{eqnarray}\label{eq:Gpecho}
    \tilde{G}^+_\echo(\omega, x,x')=\frac{T_\bh(\omega) R_\wa(\omega) e^{-2i\omega x_0}}{A_\in^\uco(\omega)}\frac{X_\bh^+(\omega, x)X_\bh^+(\omega, x')}{2i\omega}\,.
\end{eqnarray}
Here, the BH transmission coefficient $T_\bh(\omega)=1/A_\in^\bh(\omega)$ is given by Eq.~(\ref{eq:TBH}), and the additional phase factor $e^{-2i\omega x_0}$ denotes  the time delay due to interior reflection.

\begin{figure}[htbp] 
%    \centering 

    \includegraphics[width=1\textwidth]{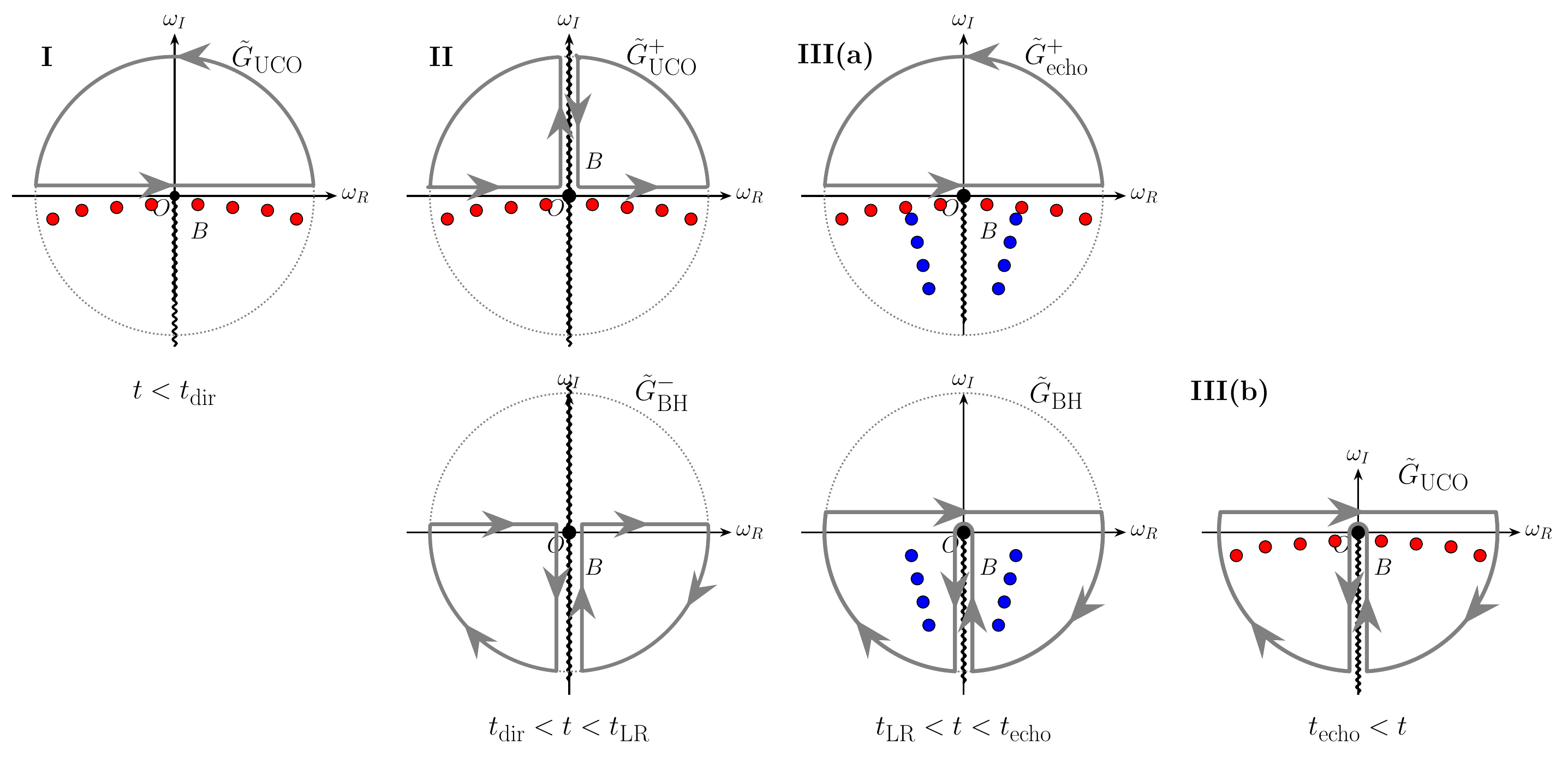}

    \caption{Similar to Fig.~\ref{fig:contour_outside}, but with a further decomposition of the UCO Green function into the BH Green function plus echo corrections in Stage III. This decomposition makes explicit the time delay $t_d$ arising from the interior reflection, and introduces the additional time scale $t_\echo(x,x')$ in Eq.~(\ref{eq:techo}) (denoted in the plot as $t_\echo$), separating Stage III into two intervals. Compared to Fig.~\ref{fig:contour_outside}, the UCO Green function in Stage III(a) is represented differently, where blue circles denote the BH QNMs.}
    \label{fig:contour_outside2} 
\end{figure}

Although the frequency-domain decomposition in Eq.~(\ref{eq:Gpucodec}) cleanly separates the BH contribution from the echo correction with the expected time delay, the roles of the BH and UCO QNMs in the actual temporal evolution remain unclear until the inverse Laplace transform is performed. In fact, to obtain a complete picture, this decomposition must be combined with the appropriate choice of integration contour--the key point emphasized in recent methodological developments for BHs. Within the framework of Fig.~\ref{fig:contour_outside}, this amounts to dividing Stage III into two intervals, as illustrated in Fig.~\ref{fig:contour_outside2}. In Stage III(a), with the decomposition of Eq.~(\ref{eq:Gpucodec}), the inverse Laplace transform is evaluated by closing the contour for $\tilde{G}_\bh$ in the lower half-plane and that for $\tilde{G}^+_\echo$ in the upper half-plane. 
Although $\tilde{G}^+_\echo$ has a complicated pole structure in the lower half-plane, each factor in Eq.~(\ref{eq:Gpecho}) is analytic in the upper half-plane, so it contributes nothing at this stage. The signal is then given entirely by the BH QNMs and tail from $\tilde{G}_\bh$.
In Stage III(b), both components are enclosed in the lower half-plane, and the response in this regime (i.e. the echoes) is again described solely by the UCO QNMs and branch cuts, exactly as in Fig.~\ref{fig:contour_outside}.

This reinterpretation of Stage III(a) in Fig.~\ref{fig:contour_outside2} completes the existing discussion in the literature. First, it provides a rigorous causality-based justification: the UCO response before $t_\echo$ is identical to that of a BH, so standard black hole spectroscopy tests only the effective BH nature up to that timescale. Second, the equality of the UCO and BH Green functions at early times implies that the BH QNM and tail expansion is equivalent to the UCO one in this stage, although, as we will discuss later, the two bases differ in reconstruction efficiency.

To make this efficiency difference explicit, we now write the time-domain Green function for UCOs. In Stage III, i.e. $t>t_\lr(x,x')$, the inverse Laplace transform in Eq.~(\ref{eq:GtInvLap}) yields a decomposition of $G_\uco(t,x,x')$ into a UCO QNM contribution and a branch-cut term. 
The branch-cut contribution is known to reproduce the BH power-law tail at late times, and for strong reflection the tail is pushed to even later times by the slowly decaying trapped modes~\cite{Rosato:2025rtr}. Hence, for the timescales of interest we just focus on the QNM part, which originates from \(\tilde{G}_\uco^+\) in Eq.~(\ref{eq:tGucopOUT}).
Because the potential is real, the QNMs appear in complex-conjugate pairs, $\pm \omega_{R,n}+i\omega_{I,n}$, and this contribution can be written as
\begin{eqnarray}
    G^\qnm_\uco(t,x,x')=\sum_n \left[B^+_n e^{-i\omega_n(t-x-x')}+B^{+*}_ne^{i\omega_n^*(t-x-x')}\right]
    =2\Re\left[\sum_{n} B^+_n e^{-i\omega_n(t-x-x')}\right],
\end{eqnarray}
where $B^+_n$ denotes the excitation factor of the mode. 
Since the excitation factors are conjugate for the two members of a QNM pair, the sum is simplified in the last expression by including only the positive-frequency representative of each pair (with $\omega_{n,R}>0$),
and $B^+_n$ is given by
\begin{eqnarray}
    B^+_n=-\frac{1}{2\omega_n}\frac{A_\out^\uco(\omega_n)}{\left.dA_\in^\uco/d\omega\right|_{\omega=\omega_n}}\,.
\end{eqnarray}
Using the relation between $A_{\rm in, out}^\uco$ and their BH counterparts in Eqs.~(\ref{eq:Rel10}) and (\ref{eq:Rel20}), together with the UCO QNM condition Eq.~(\ref{eq:UCOQNMs}), the excitation factor can be further rewritten as
\begin{eqnarray}
    B^+_n
    =-\frac{1}{2\omega_n}\frac{T_\bh(\omega_n)R_\wa(\omega_n)e^{-2i\omega_n x_0}}{\left.dA_\in^\uco/d\omega\right|_{\omega=\omega_n}}
    =-\frac{1}{2\omega_n}\frac{T_\bh(\omega_n)e^{-2i\omega_n x_\lr}}{R_\bh(\omega_n)\left.dA_\in^\uco/d\omega\right|_{\omega=\omega_n}}\,,
\end{eqnarray}
which exhibits different phase factors. Factorizing the phase factors then leads to two equivalent expressions of the QNM contribution,
\begin{eqnarray}\label{eq:GFtimeOUT1}
    G^\qnm_\uco(t,x,x')=
    2\Re\left[\sum_{n} \bar{B}^+_n e^{-i\omega_n(t-t_\lr(x,x'))}\right]
    =2\Re\left[\sum_{n} \tilde{B}^+_n e^{-i\omega_n(t-t_\echo(x,x'))}\right]\,.
\end{eqnarray}
Here, $\bar{B}^+_n$ and $\tilde{B}^+_n$ denote the mode amplitudes at the beginning of Stage III and Stage III(b) shown in Fig.~\ref{fig:contour_outside2}, respectively. Their explicit forms are given by
\begin{eqnarray}\label{eq:bBntBnout}
    \bar{B}^+_n&=&-\frac{1}{2\omega_n}\frac{T_\bh(\omega_n)}{R_\bh(\omega_n)\left.dA_\in^\uco/d\omega\right|_{\omega=\omega_n}}\approx \frac{1}{2i\omega_nt_d}\frac{T_\bh^2(\omega_n)}{R_\bh(\omega_n)}\,,\nonumber\\
    \tilde{B}^+_n&=&-\frac{1}{2\omega_n}\frac{T_\bh(\omega_n)R_\wa(\omega_n)}{\left.dA_\in^\uco/d\omega\right|_{\omega=\omega_n}}\approx \frac{1}{2i\omega_n t_d}T_\bh^2(\omega_n)R_\wa(\omega_n)\,.
\end{eqnarray}
In the large-$t_d$ limit,  the derivative for $A_\in^\uco(\omega)$ is dominated by the phase factor associated with $t_d$, i.e. $dA_\in^\uco/d\omega|_{\omega=\omega_n}\approx -it_d A_\in^\bh(\omega_n)$. The last expressions in Eq.~(\ref{eq:bBntBnout}) incorporate this approximation for both coefficients.

One immediate consequence is clear from Eq.~(\ref{eq:bBntBnout}). For QNM frequencies $\omega_{n,R}$ considerably higher than $\omega_{\textrm{RD},R}$, one has $|T_\bh(\omega_n)|\to 1$ and $|R_\bh(\omega_n)|\to 0$. As a result, the amplitudes of high-frequency modes become very large around $t_\lr(x,x')$ ($\bar{B}^+_n\gg 1$), while remain comparable to low-frequency amplitudes around $t_\echo(x,x')$. As we shall illustrate numerically in the next section, this implies that although the BH ringdown can in principle be reconstructed from the UCO QNMs, the reconstruction is inefficient for the outside source case. A large number of high-frequency modes with very accurately determined amplitudes is required to reproduce the correct interference pattern.

\subsection{Source inside the light ring}\label{sec:theory_inside}

We now turn to the case where the source lies inside the light ring ($x'\ll x_\lr\ll x$). We first examine the asymptotic behavior of the frequency-domain Green function for BH. In the large-$|\omega|$ limit,  
substituting Eqs.~(\ref{eq:SNEampp}) and (\ref{eq:SNEampm}) into Eq.~(\ref{eq:GFomega0}) yields
\begin{eqnarray}
    \tilde{G}_\bh(\omega, x,x')=\frac{1}{W_\bh(\omega)}X_\bh^+(\omega, x)X_\bh^-(\omega, x')
    \to \frac{1}{2i\omega}\frac{1}{A_\in^\bh}e^{i\omega (x-x')}\,,
\end{eqnarray}
In contrast to the outside case, the Green function involves only a single time scale and requires no decomposition~\cite{Rosato:2026moe}. Because the direct wave must cross the light-ring potential barrier to reach the observer, it is already sensitive to the barrier; we therefore denote the relevant time scale as
\begin{eqnarray}\label{eq:tlrIN}
    t_\lr(x,x')\equiv x-x'\,.
\end{eqnarray}
This is also consistent with the toy model of Ref.~\cite{Chavda:2024awq}, where the counterpart of $\tilde{G}^-_\bh(\omega,x,x')$ vanishes identically and the remaining term, which encodes the barrier, is accompanied by a step function that sets in at this new $t_\lr(x,x')$. Hence, for generic frequencies we may simply write  
$\tilde{G}_\bh(\omega, x,x')=\tilde{G}^+_\bh(\omega, x,x')$ with  
\begin{eqnarray}\label{eq:GBHm}
    \tilde{G}^+_\bh(\omega, x,x')=
    \frac{1}{A_\in^\bh(\omega)}\frac{X_\bh^+(\omega, x)X_\bh^-(\omega, x')}{2i\omega}\,.
\end{eqnarray}
The standard Leaver's contour then applies immediately for $t>t_\lr(x,x')$, and the full non-trivial response is captured by the BH QNMs and tail.

For the UCO, replacing $X_\bh^-$ by $X_\uco^-$ and again using the asymptotic forms Eqs.~(\ref{eq:SNEampp}) and (\ref{eq:SNEamppUCO}) in Eq.~(\ref{eq:GFomega0}) yields
\begin{eqnarray}
    \tilde{G}_\uco(\omega, x,x')&=&\frac{1}{W_\uco(\omega)}X_\bh^+(\omega, x)X_\uco^-(\omega, x')\nonumber\\
    &\to& \frac{1}{2i\omega}\frac{1}{A_\in^\uco(\omega)}\left[e^{i\omega (x-x')}+R_\wa e^{i\omega (x+x'-2x_0)}\right]\,.
\end{eqnarray}
Due to the interior reflection, a second time scale appears, 
\begin{eqnarray}\label{eq:tref}
    t_\re(x,x')\equiv x+x'-2x_0 = t_\lr(x,x')+t_s(x')\,,
\end{eqnarray}
where $t_s(x')=2(x'-x_0)<t_d$. 
For generic frequencies, this motivates the decomposition:  $X_\uco^-(\omega,x')=X_\bh^-(\omega,x')+A_\re^\uco X_\bh^{\prime -}(\omega,x')$, where $X_\bh^{\prime -}$ denotes the solution with outgoing boundary condition at horizon. 
The Green function is then split into two components,
\begin{eqnarray}\label{eq:GFomegaIN1}
    \tilde{G}_\uco(\omega, x,x')=\tilde{G}^+_\uco(\omega, x,x')+\tilde{G}^{++}_\uco(\omega, x,x')\,,
\end{eqnarray}
with 
\begin{eqnarray}
    \tilde{G}^+_\uco(\omega, x,x')&=&\frac{1}{A_\in^\uco(\omega)}\frac{X_\bh^+(\omega, x)X_\bh^-(\omega, x')}{2i\omega}\,,\nonumber\\
    \tilde{G}^{++}_\uco(\omega, x,x')&=&\frac{R_\wa(\omega)e^{-2i\omega x_0}}{A_\in^\uco(\omega)}\frac{X_\bh^+(\omega, x)X_\bh^{\prime -}(\omega, x')}{2i\omega}\,.
\end{eqnarray}
Compared to the outside case, the decomposition differs more markedly from the BH one:  the first component is modified by the replacement $A_\in^\bh(\omega) \to A_\in^\uco(\omega)$, and the second component--the interior reflection of the source emitted wave--is entirely absent for a BH.

\begin{figure}[htbp] 
%    \centering 

    \includegraphics[width=0.75\textwidth]{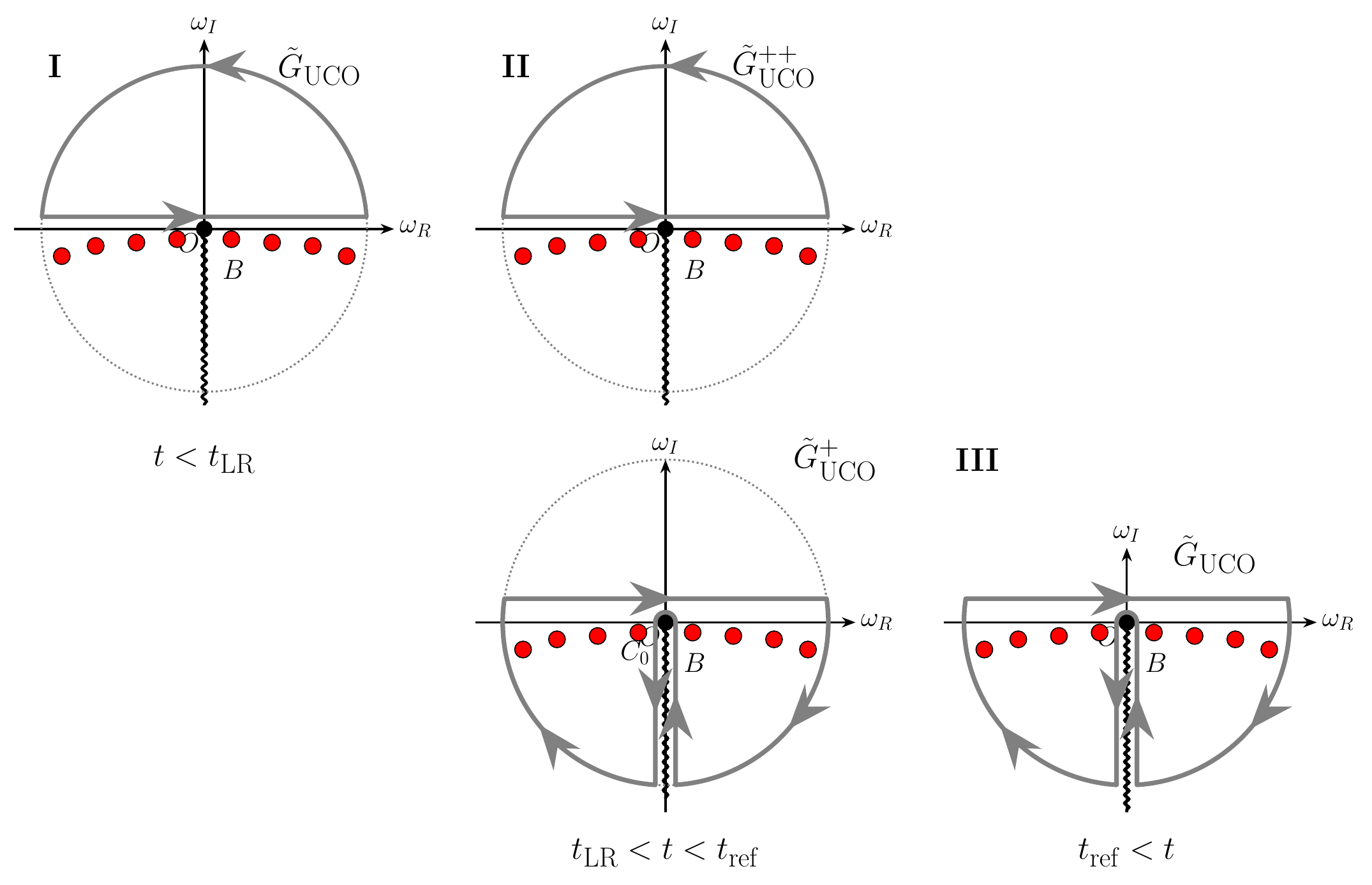}

    \caption{Decomposition of the UCO Green function and appropriate enclosing contours in the complex $\omega$-plane  for the inverse Laplace transform at different evolution stages, for a source inside the light ring. The three stages are separated by the time scales $t_\lr(x,x')$ and $t_\re(x,x')$ (see Eqs.~(\ref{eq:tlrIN}) and (\ref{eq:tref})), denoted in the plot simply as $t_\lr$ and $t_\re$. Red circles mark the UCO QNMs; $B$ denotes the branch cuts.}
    \label{fig:contour_inside} 
\end{figure}

Substituting Eq.~(\ref{eq:GFomegaIN1}) into the inverse Laplace transform Eq.~(\ref{eq:GtInvLap}) and imposing causality leads to the contour choices illustrated in Fig.~\ref{fig:contour_inside}. 
In Stage I, $\tilde{G}_\uco$ is again analytic in the upper half-plane, so there is no signal here. In Stage II, the contour for $\tilde{G}^+_\uco$ is closed in the lower half-plane, while that for $\tilde{G}^{++}_\uco$ is closed in the upper half-plane; the latter is analytic there and contributes nothing. The signal is therefore given by the UCO QNMs and tail from $\tilde{G}^+_\uco$ alone. In Stage III, both components are evaluated by closing their contours in the lower half-plane, and the response is again governed by the UCO QNMs and tail contributions. Thus, even at this level, ``quantum BH seismology'' already exhibits a richer structure than in the outside case.

As for the outside case, one may wonder why the early signal before $t_\re(x,x')$ should reproduce the BH ringdown. The resolution is the same: the UCO amplitude ratio can be expanded as the BH part plus echo corrections. Using Eq.~(\ref{eq:Rel11}) for $1/A_\in^\uco(\omega)$ away from the QNM poles, $\tilde{G}^+_\uco(\omega, x,x')$ in Eq.~(\ref{eq:GFomegaIN1}) decomposes as
\begin{eqnarray}\label{eq:GFomegaIN2}
    \tilde{G}^+_\uco(\omega, x,x')=\tilde{G}^+_\bh(\omega, x,x')+\tilde{G}^+_\echo(\omega, x,x')\,,
\end{eqnarray}
with $\tilde{G}^+_\bh(\omega, x,x')$ given in Eq.~(\ref{eq:GBHm}) and 
\begin{eqnarray}
    \tilde{G}^+_\echo(\omega, x,x')=
    \frac{R_\bh(\omega) R_\wa(\omega) e^{i\omega t_d}}{A^\uco_\in(\omega)}\frac{X_\bh^+(\omega, x)X_\bh^-(\omega, x')}{2i\omega}\,.
    %\frac{T_\bh R_\bh R_\wa e^{i\omega t_d}}{1-R_\bh R_\wa e^{i\omega t_d}}\frac{X_\bh^+(\omega, x)X_\bh^-(\omega, x')}{2i\omega}\,.
\end{eqnarray}
In the large-$|\omega|$ limit, the echo term is governed by the phase factor $e^{i\omega(x-x'+t_d)}$,  which defines a third time scale associated with the interior reflection:
\begin{eqnarray}\label{eq:techoIN}
    t_\echo(x,x')\equiv x-x'+t_d=t_\lr(x,x')+t_d\,.
\end{eqnarray}

\begin{figure}[htbp] 
%    \centering        
    \includegraphics[width=1\textwidth]{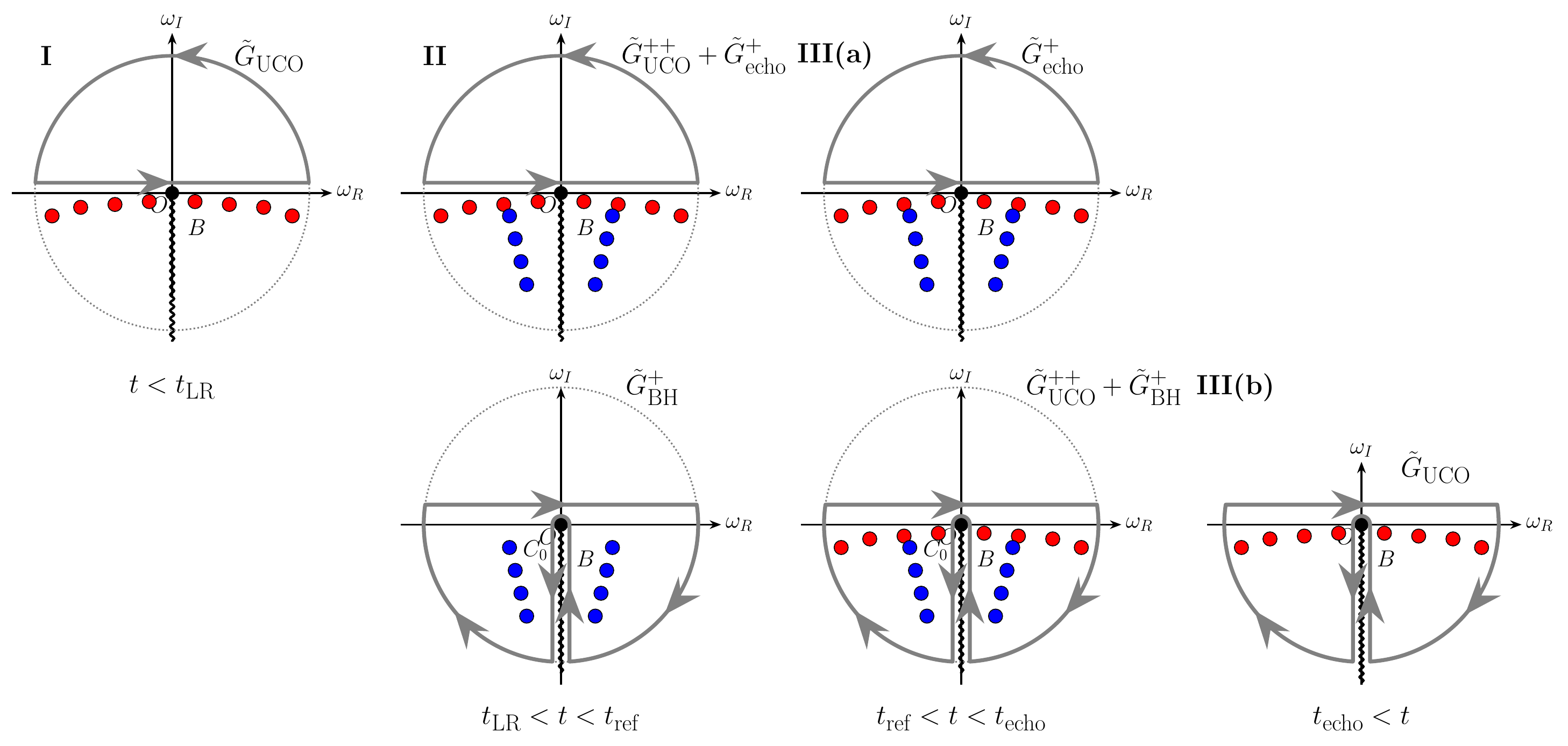}     
    \caption{Similar to Fig.~\ref{fig:contour_inside}, but with a further decomposition of $\tilde{G}_\uco^+$ into its BH counterpart plus echo corrections, starting from Stage II. By making explicit the time delay $t_d$ arising from interior reflection, this decomposition again introduces the additional time scale $t_\echo(x,x')$ in Eq.~(\ref{eq:techoIN}) (denoted in the plot as $t_\echo$) to split Stage III into two intervals. Compared to Fig.~\ref{fig:contour_inside}, Stage II and Stage III(a) are now represented differently, where blue circles mark the BH QNMs.}
    \label{fig:contour_inside2} 
\end{figure}

Combining this further decomposition with the appropriate contour choices splits Stage III of Fig.~\ref{fig:contour_inside} into two intervals, as shown in Fig.~\ref{fig:contour_inside2}. 
Because Eq.~(\ref{eq:GFomegaIN2}) also affects Stage II, the contour assignment for $t_\lr(x,x') < t < t_\re(x,x')$ becomes even cleaner: only $\tilde{G}^+_\bh$ is enclosed in the lower half-plane, yielding exactly the BH QNMs and tail. Both $\tilde{G}^+_\echo$ and $\tilde{G}^{++}_\uco$ are enclosed in the upper half-plane and, being analytic there, give zero contribution.  
In Stage III(a), the contours for $\tilde{G}^+_\bh$ and $\tilde{G}^{++}_\uco$ are closed in the lower half-plane, giving the BH QNMs and tail together with UCO QNM contributions from $\tilde{G}^{++}_\uco$; $\tilde{G}^+_\echo$ remains in the upper half-plane and contributes nothing.  
In Stage III(b), all components are enclosed in the lower half-plane, and the signal is governed by the full UCO QNM and tail spectrum.  

Two key differences from the outside case emerge. First, echoes can appear at much earlier times if $t_s(x') \ll t_d$, depending on how close the source is to the interior surface. Second, because $\tilde{G}^+_\uco$ and $\tilde{G}^{++}_\uco$ are controlled by different time scales, the signal is an interference of two echo trains, resulting in a more complicated time-domain structure. This feature already appears in the two delta-function toy model of Ref.~\cite{Hui:2019aox}.

Although the conceptual lesson regarding the equivalence of BH and UCO QNM expansions at early times remains the same as for the outside case, the practical reconstruction efficiency differs markedly. To see this, we write the time-domain Green function explicitly. 
In Stage II, i.e. $t_\lr(x,x')<t<t_\re(x,x')$, only $\tilde{G}^+_\uco$ contributes in the lower half-plane, and the UCO QNM sum reads
\begin{eqnarray}\label{eq:GFtimeIN1}
    G_\uco^{\qnm}(t,x,x')=2\Re\left[\sum_{n} \bar{B}^+_n e^{-i\omega_n(t-t_\lr(x,x'))}\right]\,.
\end{eqnarray}
The excitation factor for the positive-frequency modes is given by
\begin{eqnarray}\label{eq:Bnm}
    \bar{B}^+_n=-\frac{1}{2\omega_n}\frac{1}{\left.dA_\in^\uco/d\omega\right|_{\omega=\omega_n}}
    \approx \frac{1}{2i\omega_n t_d}T_\bh(\omega_n)\,,
\end{eqnarray}
which represents the mode amplitude at $t=t_\lr(x,x')$. 
Unlike the outside case, where $\bar{B}^+_n\propto T^2_\bh(\omega_n)/R_\bh(\omega_n)$ (Eq.~(\ref{eq:bBntBnout})) grows exponentially at high frequencies, here $\bar{B}_n^+$ is proportional to $T_\bh(\omega_n)$ and increases far less rapidly. Consequently, a relatively small number of UCO QNMs can efficiently reconstruct the BH ringdown in this case. We will verify this prediction numerically in the next section.

In Stage III(a), both $\tilde{G}^+_\uco$ and $\tilde{G}^{++}_\uco$ contribute in the lower half-plane, and the QNM expansion becomes 
\begin{eqnarray}\label{eq:GFtimeIN2}
    G_\uco^{\qnm}(t,x,x')=2\Re\left[\sum_{n} \bar{B}^+_n e^{-i\omega_n(t-t_\lr(x,x'))}\right]+2\Re\left[\sum_{n} \tilde{B}^{++}_n e^{-i\omega_n(t-t_\re(x,x'))}\right]\,.
\end{eqnarray}
The first part, involving the same $B_n^+$ as in Eq.~(\ref{eq:Bnm}), reproduces the BH QNMs and tail.  
The second part, with the excitation factor 
\begin{eqnarray}\label{eq:tBnp}
    \tilde{B}^{++}_n=\bar{B}^+_n R_\wa(\omega_n)
    \approx \frac{1}{2i\omega_n t_d}T_\bh(\omega_n)R_\wa(\omega_n)\,,
\end{eqnarray}
corresponds to the first interior reflection at $t = t_\re(x,x')$. 
In Stage III(b), both sets of QNMs again contribute. To make the echo delay $t_d$ explicit, we rewrite $\bar{B}_n^+$ by factoring out the phase $e^{i\omega_n t_d}$ with the help of the UCO QNM condition in Eq.~(\ref{eq:UCOQNMs}).  
The Green function in Eq.~(\ref{eq:GFtimeIN2}) can then be expressed as
\begin{eqnarray}\label{eq:GFtimeIN3}
    G_\uco^{\qnm}(t,x,x')=2\Re\left[\sum_{n} \tilde{B}^+_n e^{-i\omega_n(t-t_\echo(x,x'))}\right]+2\Re\left[\sum_{n} \tilde{B}^{++}_n e^{-i\omega_n(t-t_\re(x,x'))}\right]\,,
\end{eqnarray}
where
\begin{eqnarray}\label{eq:tBnm}
    \tilde{B}^+_n=-\frac{1}{2\omega_n}\frac{R_\bh(\omega_n)R_\wa(\omega_n)}{\left.dA_\in^\uco/d\omega\right|_{\omega=\omega_n}}
    \approx \frac{1}{2i\omega_n t_d}T_\bh(\omega_n)R_\bh(\omega_n)R_\wa(\omega_n),
\end{eqnarray}
denotes the mode amplitudes at $t=t_\echo(x,x')$. 
Compared to the outside case, the coefficients $\tilde{B}_n^+$ and $\tilde{B}_n^{++}$ in Eqs.~(\ref{eq:tBnp}) and (\ref{eq:tBnm}) are less suppressed by $T_\bh(\omega_n)$ at low frequencies, while $\tilde{B}_n^+$ is further suppressed by $R_\bh(\omega_n)$ at high frequencies.  
This indicates that low frequency trapped modes play a more prominent role in the inside case.

\section{Numerical validations}
\label{sec:validaiton}

In this section we numerically verify the theoretical framework developed in Sec.~\ref{sec:theory}.  
To that end, we perform direct time-domain simulations of the waveform $\psi(t,x)$ to approximate the Green function using Eq.~(\ref{eq:EOMtime}); these simulations serve as a reference against which the decomposition schemes illustrated in Figs.~\ref{fig:contour_outside}- \ref{fig:contour_inside2} can be tested.  
Agreement between the simulations and the different decomposition then provides a unified picture of the UCO time-domain Green function and its connection to the various sets of QNMs.

We consider axial gravitational perturbations governed by the Regge-Wheeler equation in this work. For this case, the potential in Eq.~(\ref{eq:EOMtime}) takes the form 
\begin{equation}
    V(r) = \left( 1 - \frac{2M}{r} \right) \left( \frac{\ell(\ell+1)}{r^2} - \frac{6M}{r^3} \right).
\end{equation}
We focus on the $\ell=2$ perturbations; the corresponding fundamental ($n=0$) QNM frequency for a BH is $M\omega_{{\rm RD},R}=0.37$. 
The conventional tortoise coordinate $x = r + 2M \ln\left(r/2M - 1\right)$ is used throughout, with the light ring position $x_\lr\approx 1.6M$.

To approximate the Green function as closely as possible from the time evolution of Eq.~(\ref{eq:EOMtime}), we choose initial data with a vanishing field value and a narrow Gaussian profile for the time derivative,
\begin{equation}
    \psi(t=0, x) = 0, \quad \left.\frac{\partial\psi(t,x)}{\partial t}\right|_{t=0} = \frac{1}{\sqrt{2\pi}\sigma} \exp\left[ -\frac{(x - x_s)^2}{2\sigma^2} \right],
\end{equation}
where $t_0=0$. 
The width $\sigma$ of the Gaussian is chosen sufficiently narrow so that the waveform closely approximates the Green function and the QNM reconstruction discussed earlier can be directly compared with the numerical waveform. Specifically, we place both the observer and the source far from the potential barrier: the observer is located at $x_\obs = 60M$, and the Gaussian profile is centered at $x_s = 40M$ for the outside case and $x_s = -30M$ for the inside case. With the choice $\sigma = 0.01M$, we then identify $G(t,x_\obs,x_s) \approx \psi(t,x_\obs)$ (see Appendix~\ref{app:initialcond} for further details on the narrow-Gaussian approximation for the Green function).
Eq.~(\ref{eq:EOMtime}) is evolved with a second-order finite-difference scheme, using numerical step sizes $\Delta x = 0.001M$ and $\Delta t = 0.0005M$. The boundary conditions are chosen as follows. The outer boundary is placed at a large value $x_{\rm max}= 2000M$, where the first-order Sommerfeld condition $(\partial_t + \partial_x)\psi(t,x) = 0$ is imposed, allowing outgoing waves to leave the computational domain without spurious reflections. The final evolution time is set to $t_{\rm max} = 4000M$.

\begin{figure}[htbp] 
    \centering 
    \includegraphics[width=0.6\textwidth]{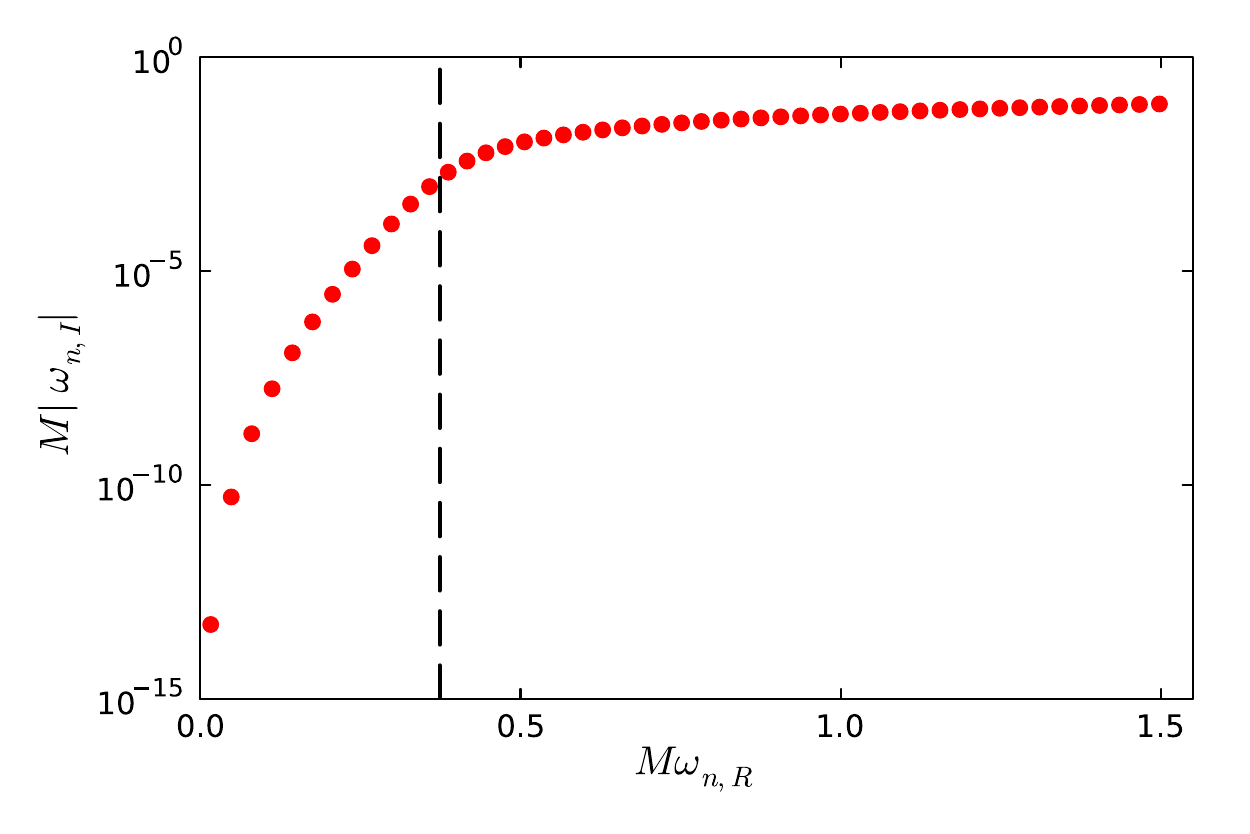}
    \caption{UCO QNM frequencies for the benchmark model with $x_0=-100M$ and $R_\wa(\omega)=1$. The low-frequency modes accumulate close to the real axis and correspond to the long-lived trapped branch responsible for late-time echoes, while the high-frequency modes form a separate branch with much larger damping. The vertical dashed line marks the fundamental BH ringdown frequency, $M\omega_{{\rm RD},R}\simeq 0.37$.}
    \label{fig:UCOQNMs} 
\end{figure}

The inner boundary is the reflective surface at $x = x_0$, with the reflection coefficient $R_\wa(\omega)$ defined in Eq.~(\ref{eq:Aucoref}). For the numerical validation, 
%we consider a benchmark model with
we choose a single benchmark model to illustrate the key features: 
a highly compact object with $x_0 = -100M$ (i.e. a round-trip delay $t_d\approx 203.2M$), and a perfectly reflecting surface, $R_\wa(\omega) = 1$, which corresponds to a Neumann boundary condition $\partial_x \psi(t,x_0) = 0$ in the time domain.
%$x_0 = -100M$ and $R_\wa(\omega) = 1$ for the interior surface,  corresponding to a strongly reflecting, highly compact object with $t_d\approx 203.2M$. 
The QNM reconstruction requires an accurate computation of the asymptotic amplitudes at complex frequencies. To this end we employ the recent package \texttt{GeneralizedSasakiNakamura.jl}~\cite{Lo:2023fvv,Lo:2025njp}, which computes the Kerr BH amplitudes in the Sasaki-Nakamura formalism, and we take the spinless limit to extract the Schwarzschild amplitudes.

Figure~\ref{fig:UCOQNMs} displays the real and imaginary frequencies of the UCO QNMs for this benchmark, obtained by numerically solving Eq.~(\ref{eq:UCOQNMs2}). 
The phase $\phi_n$ there varies only mildly with $n$ compared to the $2\pi n$ term, so the real part of the QNMs scales nearly linearly with $n$.  The imaginary part, however, behaves very differently at low and high frequencies. Low-frequency (trapped) modes, with $\omega_{n,R}<\omega_{{\rm RD},R}$, have small imaginary parts. In this regime, $R_\bh(\omega_n)R_\wa(\omega_n) \approx R_\bh(\omega_{n,R})R_\wa(\omega_{n,R})$, and the second equation in Eq.~(\ref{eq:UCOQNMs2}) reduces to 
\begin{align}
    t_d \omega_{n,I} \approx \ln \left| R_\bh(\omega_{n,R}) R_\wa(\omega_{n,R}) \right|\,.
\end{align}
Since $R_\bh(\omega_n)$ quickly approaches unity as $\omega_n$ decreases, $\omega_{n,I}$ is proportional to its small deviation from unity. These long-lived modes are responsible for the late-time echo signal. At high frequencies, on the other hand, $\omega_{n,I}$ increases linearly with $n$ as $\omega_{n,R}$; these modes decay more rapidly and therefore play a more important role in the early-time signal~\cite{Rosato:2025byu,Rosato:2025lxb}.

\subsection{Source outside the light ring}

\begin{figure}[htbp]
    \centering
    \includegraphics[width=1\textwidth]{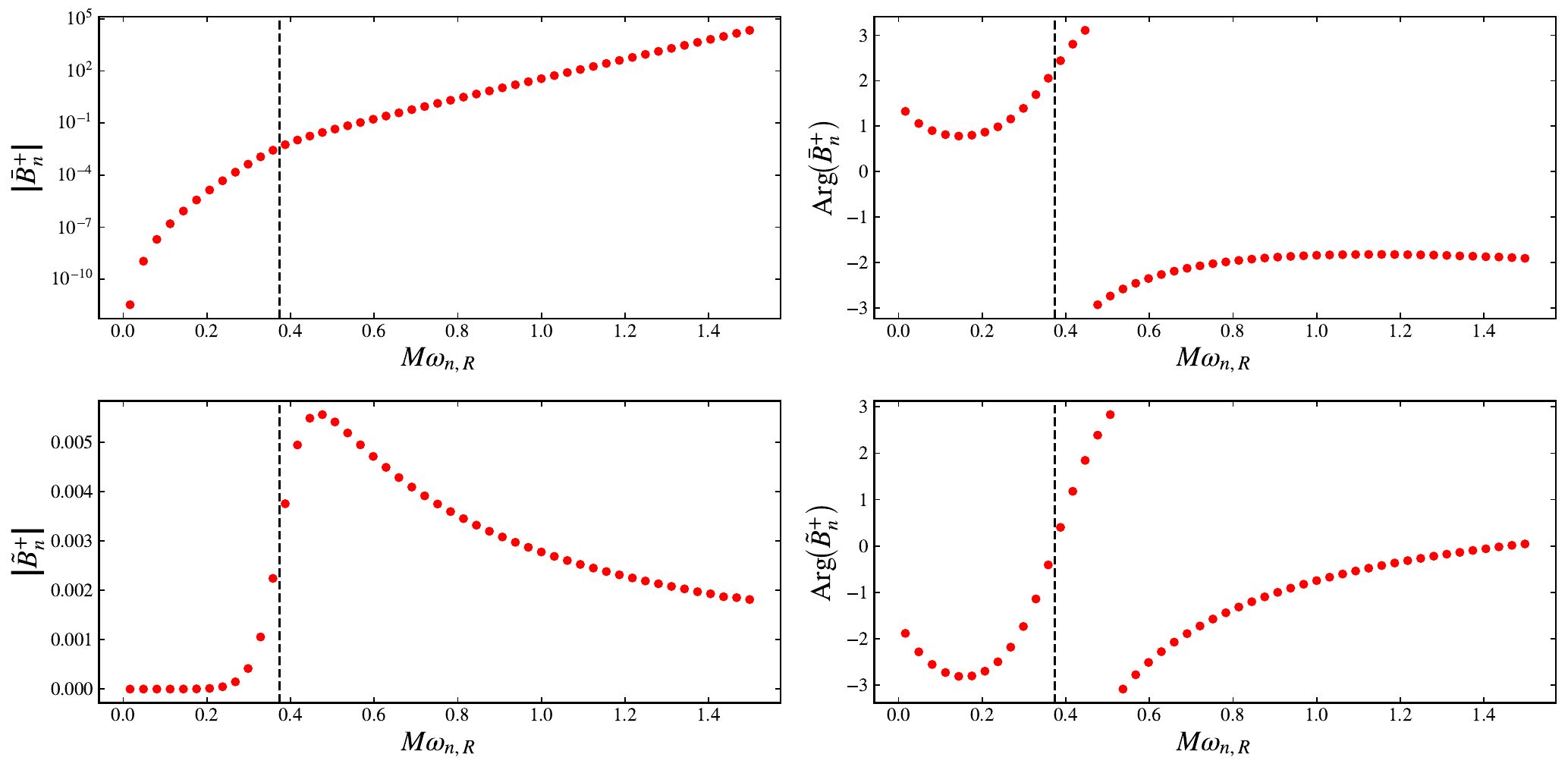}
    \caption{Modulus and phase of the excitation factors $\bar{B}^+_n$ (top) and $\tilde{B}^+_n$ (bottom) for the UCO QNMs in the outside case, computed for the benchmark model $x_0=-100M$ and $R_\wa(\omega)=1$. The vertical dashed line marks the fundamental BH ringdown frequency $M\omega_{{\rm RD},R}\simeq 0.37$.}
    \label{fig:UCOBnoutside}
\end{figure}

To construct the QNM contribution, the key inputs besides the QNM frequencies are their excitation factors. Since the time domain Green function is obtained from the real part of the QNM sum, both the modulus and phase of the excitation factors matter. Figure~\ref{fig:UCOBnoutside} displays these properties for the two sets of excitation factors $\bar{B}^+_n$ and $\tilde{B}^+_n$ relevant to the outside case, as given in Eq.~(\ref{eq:bBntBnout}).\footnote{We use the exact definitions with the derivative of the asymptotic amplitude. The results are also in excellent agreement with the large $t_d$ approximation for our benchmark model.} 

At low frequencies the magnitudes of the two sets are comparable, and both are strongly suppressed by the small imaginary parts of the QNMs. At high frequencies, on the other hand, $|\tilde{B}^+_n|$ decays as $1/|\omega_n|$, whereas $|\bar{B}^+_n|$ increases exponentially.\footnote{The counterpart of $\bar{B}^+_n$ was computed in Ref.~\cite{Rosato:2025lxb}, where the low-frequency suppression was highlighted; the exponential growth at high frequencies, however, has not been discussed.} This behavior can also be understood from a different perspective: $\bar{B}^+_n$ and $\tilde{B}^+_n$ are linked by a temporal evolution over the timescale $t_d$. Consequently,  the amplitudes of low-frequency, long-lived modes vary little, while those of high-frequency, fast-decaying modes become exponentially large at early times. 
The phases of the two coefficients differ by $\phi_n$, as given in Eq.~(\ref{eq:UCOQNMs2}). They vary significantly around $\omega_{\textrm{RD},R}$ and approach asymptotic values at high frequencies.

\begin{figure}[htbp]
    \centering
    \includegraphics[width=1\textwidth]{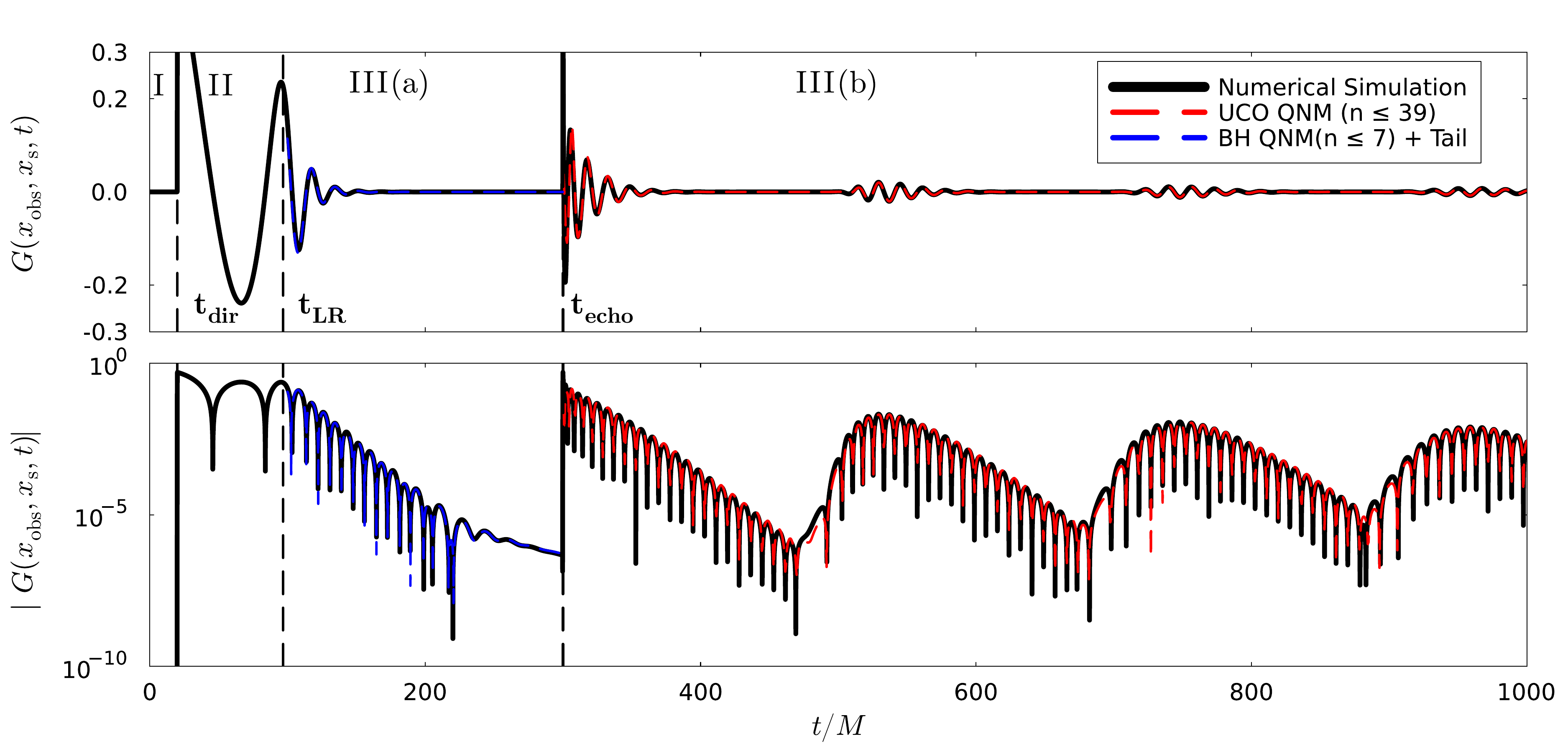}
    \caption{Temporal evolution of the UCO Green function $G_\uco(x_\obs,x_s,t)$ for a source located outside the light ring. Here, the observer is placed at $x_\obs= 60M$, and the central position of the initial velocity Gaussian profile is $x_s = 40M$. The interior reflection parameters are $x_0=-100M$ and $R_\wa(\omega)=1$. The black solid line represents the numerical simulation. The blue dashed line represents the BH QNM and tail reconstruction for modes $n=0,...7$, while the red dashed line corresponds to the UCO QNM reconstruction with modes $n=0,...,39$. Vertical dashed lines mark the three characteristic time scales $t_\dir(x_\obs,x_s)=20M$, $t_\lr(x_\obs,x_s)=96.8M$, and $t_\echo(x_\obs,x_s)= 300M$ (see Eqs.~(\ref{eq:tdir}), (\ref{eq:tlr}), and (\ref{eq:techo})), denoted in the plot simply as $t_\dir$, $t_\lr$, and $t_\echo$. The top panel displays the results on a linear scale, and the bottom panel shows the absolute value on a logarithmic scale.}
    \label{fig:FullCompOUT}
\end{figure}

We now compare the numerical simulation of the time-domain Green function with the UCO QNM reconstructions that follow from the decomposition of Sec.~\ref{sec:theory_outside}. 
Fig.~\ref{fig:FullCompOUT} shows the full time-domain evolution, incorporating all stages depicted in Fig.~\ref{fig:contour_outside}. Because the initial Gaussian profile is sufficiently narrow, the relevant time scales are well approximated by replacing the source location $x'$ with the central position $x_s$ of the Gaussian, and the QNM reconstruction is expected to match the numerical results closely except in narrow intervals of width $\sim\mathcal{O}(\sigma)$ around the characteristic time scales (see Appendix~\ref{app:initialcond} for details). 
The evolution in Stage II is insensitive to the light-ring potential barrier and is exactly the same as for a BH, dominated by $\tilde{G}^-_\bh$. The QNM reconstruction becomes relevant only in Stage III, where the influence of the curved spacetime becomes important.

As emphasized in Sec.~\ref{sec:theory_outside}, the UCO QNM reconstruction--given by the first equality in Eq.~(\ref{eq:GFtimeOUT1}) with the excitation coefficient $\bar{B}^+_n$ from Eq.~(\ref{eq:bBntBnout})--in principle applies from the start of Stage III. In particular, as illustrated by Fig.~\ref{fig:contour_outside2}, it must reproduce the exact BH ringdown throughout Stage III(a). However, Fig.~\ref{fig:UCOBnoutside} shows that $\bar{B}^+_n$ grows exponentially at high frequencies, making this representation highly inefficient in practice. That is, an extremely large number of high-frequency modes, with precisely determined amplitudes, would be required to achieve destructive interference and reproduce the correct pattern in this stage.\footnote{For a related problem, Ref.~\cite{Oshita:2025ibu} has explicitly demonstrated that the BH ringdown can be reconstructed from a completely different set of QNMs, at the cost of requiring a very large number of modes.} 
In the inside case we will present a successful reconstruction of the BH ringdown using UCO QNMs. Here, for the outside case, we therefore employ the BH QNM and tail reconstruction throughout Stage III(a), adopting the BH QNM excitation factors from Ref.~\cite{Lo:2025njp} and the Price power-law tail~\cite{Price:1971fb}.  We then switch to the UCO QNM expansion of Eq.~(\ref{eq:GFtimeOUT1}) (with the excitation factor $\tilde{B}^+_n$ from Eq.~(\ref{eq:bBntBnout})) in Stage III(b). The resulting QNM reconstructions agree well with the numerical simulation for all $t>t_\lr(x_\obs,x_s)$. 

\begin{figure}[htbp]
    \centering
    \includegraphics[width=0.95\textwidth]{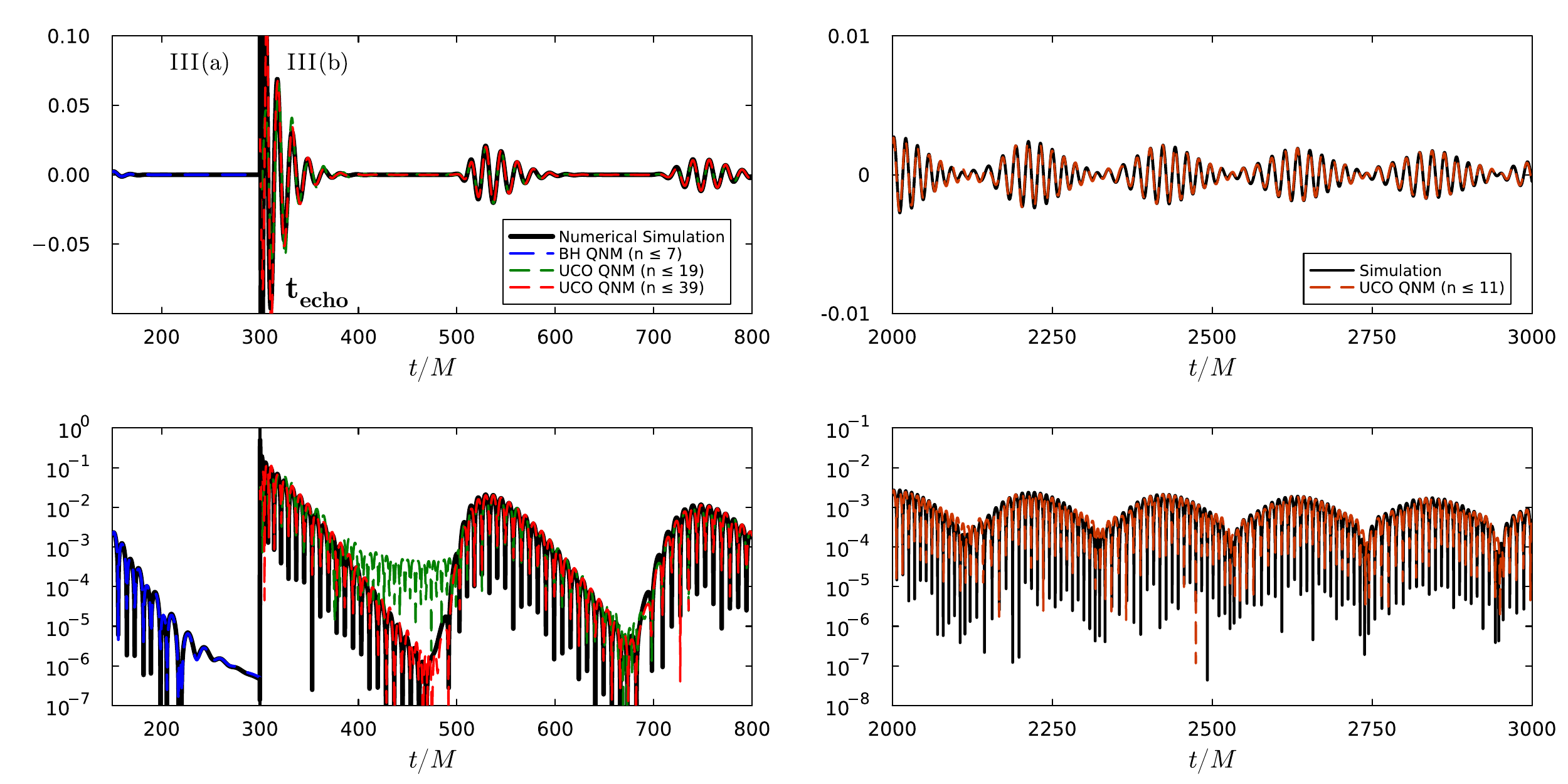}
    \caption{Detailed comparison of the UCO Green function $G_\uco(x_\obs,x_s,t)$ for the outside case. Left: temporal evolution across the transition from Stage III(a) to Stage III(b) around $t_\echo(x_\obs,x_s)$. The blue dashed line shows the BH QNM reconstruction (modes $n=0,...7$), while the red and green dashed lines correspond to the UCO QNM reconstruction with modes $n=0,...,39$ and $n=0,...,19$, respectively. Right: late time evolution at $t\gg t_\echo(x_\obs,x_s)$. The red dashed line represents the UCO QNM reconstruction using only the long-lived modes $n=0,...,11$. The top and bottom rows display the results on linear and logarithmic scales, respectively.}
    \label{fig:FullCompOUT_details}
\end{figure}

For a closer inspection, the left column of Fig.~\ref{fig:FullCompOUT_details} zooms in on the transition from Stage III(a) to Stage III(b).  Around  $t_\echo(x_\obs,x_s)$, the Green function undergoes an abrupt change from the rapidly decaying BH QNMs and tail to the first echo pulse. After this time, the UCO QNM reconstruction becomes efficient, and we compare its performance for different numbers of modes. Increasing the number of high-frequency modes generally improves the reconstruction. More importantly, these modes are essential for capturing both the deep interference dips between successive echo pulses and the sharp rise immediately after $t_\echo(x_\obs,x_s)$. 
Specifically, the first 20 modes already reproduce the echo signal well, except for the deep dip between the first and second pulses. Adding another 20 high frequency modes accurately recovers this dip. Reconstructing the early-time BH ringdown and the even sharper transition to the first echo then likely require several dozen additional modes, once again highlighting the inefficiency of the UCO QNM representation in this regime.
At late times, the high-frequency modes decay rapidly, and the low-frequency trapped modes alone suffice to give an accurate reconstruction, as shown in the right column of Fig.~\ref{fig:FullCompOUT_details}. These results are consistent with findings in the existing literature~\cite{Rosato:2025byu,Rosato:2025lxb, OuldElHadj:2026zpp}.

\subsection{Source inside the light ring}

\begin{figure}[htbp]
    \centering
    \includegraphics[width=1\textwidth]{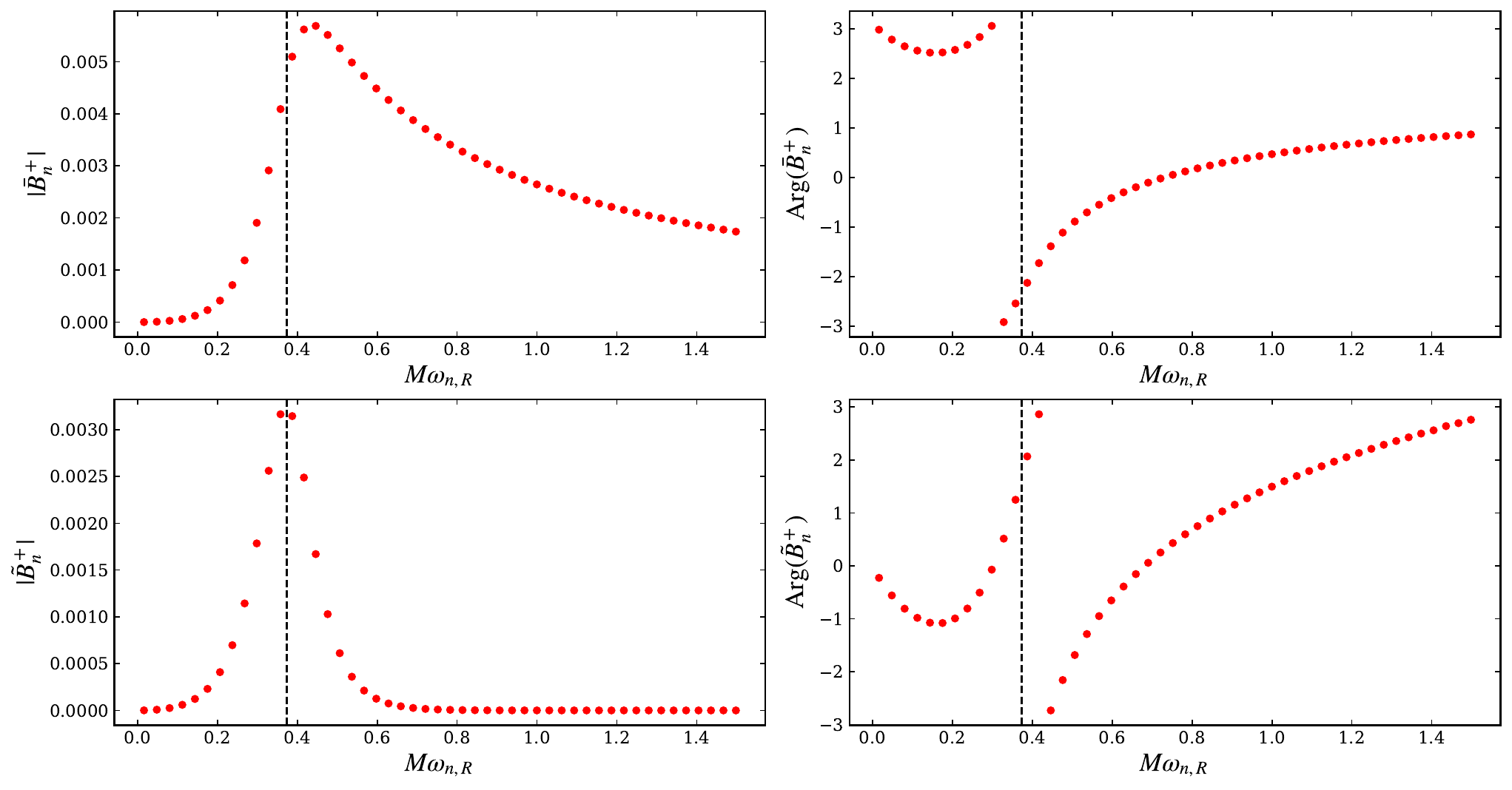}
    \caption{Modulus and phase of the excitation factors $\bar{B}_n^+$ (top) and $\tilde{B}_n^+$ (bottom) for the UCO QNMs in the inside case, computed for the benchmark model $x_0=-100M$ and $R_\wa(\omega)=1$. The vertical dashed line marks the fundamental BH ringdown frequency $M\omega_{{\rm RD},R}\simeq 0.37$.}
    \label{fig:UCOBn1inside}
\end{figure}

We now turn to the complementary case where the source is located inside the light ring. We first examine the excitation factors relevant for the UCO QNM reconstruction. For the benchmark model, $R_\wa(\omega)=1$ implies $\tilde{B}_n^{++}=\bar{B}_n^+$ from Eq.~(\ref{eq:tBnp}), so only two distinct sets of excitation factors remain.   Figure~\ref{fig:UCOBn1inside} displays the modulus and phase of $\bar{B}_n^+$ and $\tilde{B}_n^+$, as given in Eqs.~(\ref{eq:Bnm}) and (\ref{eq:tBnm}). In contrast to the outside case, the magnitudes of the two sets are quite comparable at all frequencies; this reflects the fact that the initial condition for the inside case contains far less high-frequency content. At high frequencies, $\tilde{B}_n^+$ is more suppressed than $\bar{B}_n^+$  due to the additional $R_\bh(\omega_n)$ dependence. 
The phases again differ by $\phi_n$ and exhibit a variation trend similar to that of the outside case.

\begin{figure}[htbp]
    \centering
    \includegraphics[width=1\textwidth]{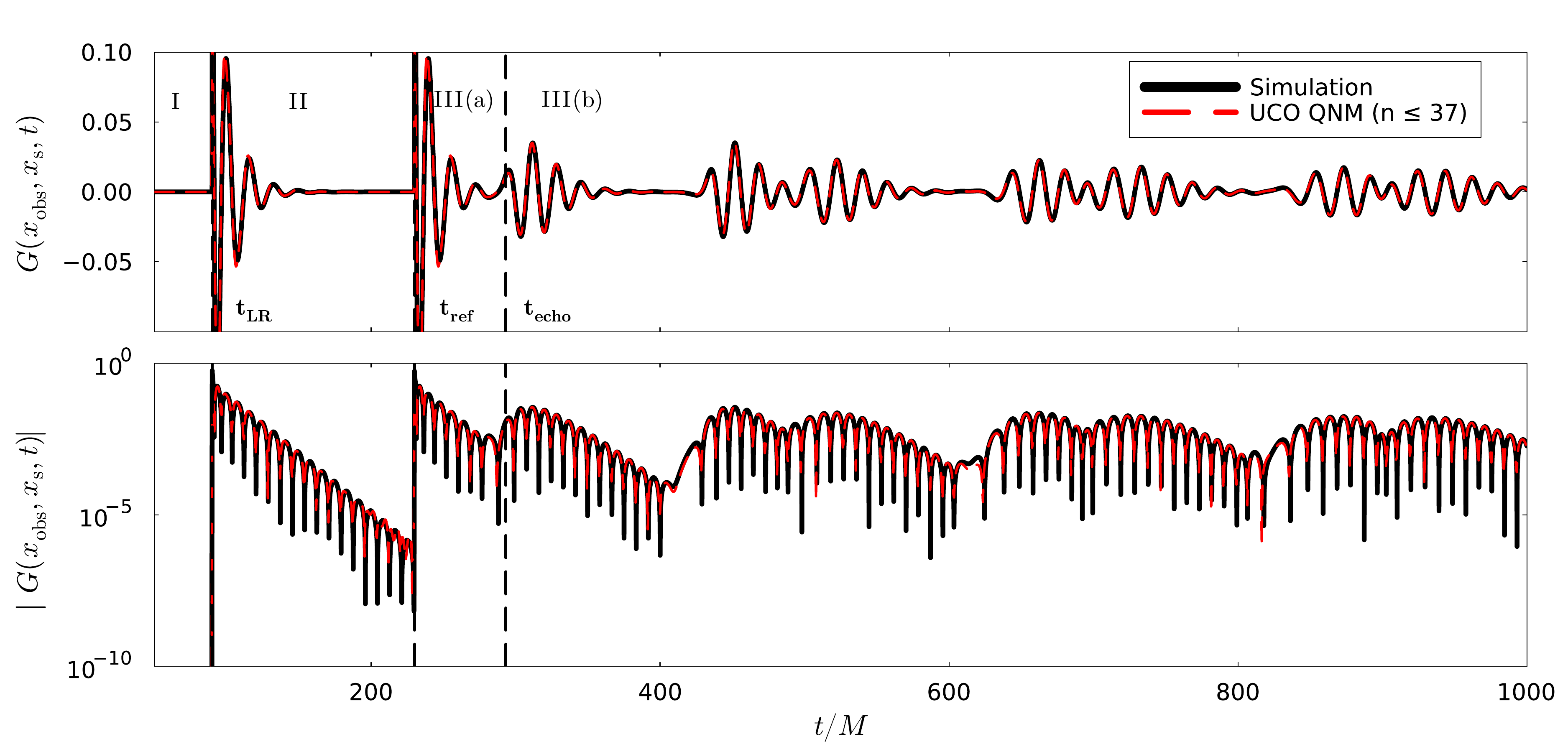}\\    
    \caption{Temporal evolution of the time-domain Green function for a source located inside the light ring. The observer is placed at $x_\obs=60M$, the central position of the initial velocity Gaussian profile is $x_s=-30M$, and the interior reflection parameters are $x_0=-100M$ and $R_\wa(\omega)=1$. The black solid line represents the numerical simulation.  The red dashed line corresponds to the UCO QNM reconstruction with modes $n=0,...,37$. Vertical dashed lines mark the three characteristic time scales $t_\lr(x_\obs,x_s)=90M$, $t_\re(x_\obs,x_s) = 230M$, $t_\echo(x_\obs,x_s) = 293.2M$ (see Eqs.~(\ref{eq:tlrIN}), (\ref{eq:tref}), and (\ref{eq:techoIN})), denoted in the plot simply as $t_\lr$, $t_\re$, and $t_\echo$. The top panel displays the results on a linear scale, and the bottom panel shows the absolute value on a logarithmic scale.}
    \label{fig:FullCompIN}
\end{figure}

Figure~\ref{fig:FullCompIN} shows the full time-domain comparison of the Green function for the inside case, incorporating all stages depicted in Fig.~\ref{fig:contour_inside}. %The corresponding time scales are marked by vertical lines. 
As expected for an inside source, the curved-spacetime influence is imprinted on the signal from the very beginning. The UCO QNM reconstruction follows the decomposition derived in Sec.~\ref{sec:theory_inside} and illustrated in Fig.~\ref{fig:contour_inside}: in Stage II we use Eq.~(\ref{eq:GFtimeIN1}) with the first set of excitation factors $\bar{B}^+_n$, and in Stage III we add the contribution of the second set $\tilde{B}^{++}_n$ according to Eq.~(\ref{eq:GFtimeIN2}). 
Interestingly, with around 40 modes the UCO QNM reconstruction accurately reproduces the full non-trivial temporal evolution from the very beginning of Stage II. This agreement has two important implications.

First, it numerically verifies the equivalence of the two representations of Stage II shown in Figs.~\ref{fig:contour_inside} and~\ref{fig:contour_inside2}: a drastically different set of UCO QNMs can indeed reproduce the exact BH ringdown signal at early times. Owing to the low-frequency dominance in this case, the UCO QNMs do provide a good expansion basis, albeit less efficient than the BH QNM basis. 
Second, it demonstrates that the UCO QNMs can reproduce a more complicated echo morphology than the simple quasiperiodic decaying pulses seen in the outside case. Specifically, the late echo signal for the inside case arises from the interference of two echo trains. The first train, which is absent in the outside case and appears only after $t_\re$, is associated with $\tilde{G}^{++}_\uco$ and described by the excitation factor $\tilde{B}_n^{++}$. The second train emerges a time $t_d$ after the initial BH ringdown, exactly as in the outside case, and is described by the excitation factors $\tilde{B}_n^+$.

\begin{figure}[htbp] 
    \centering 
    \includegraphics[width=0.95\textwidth]{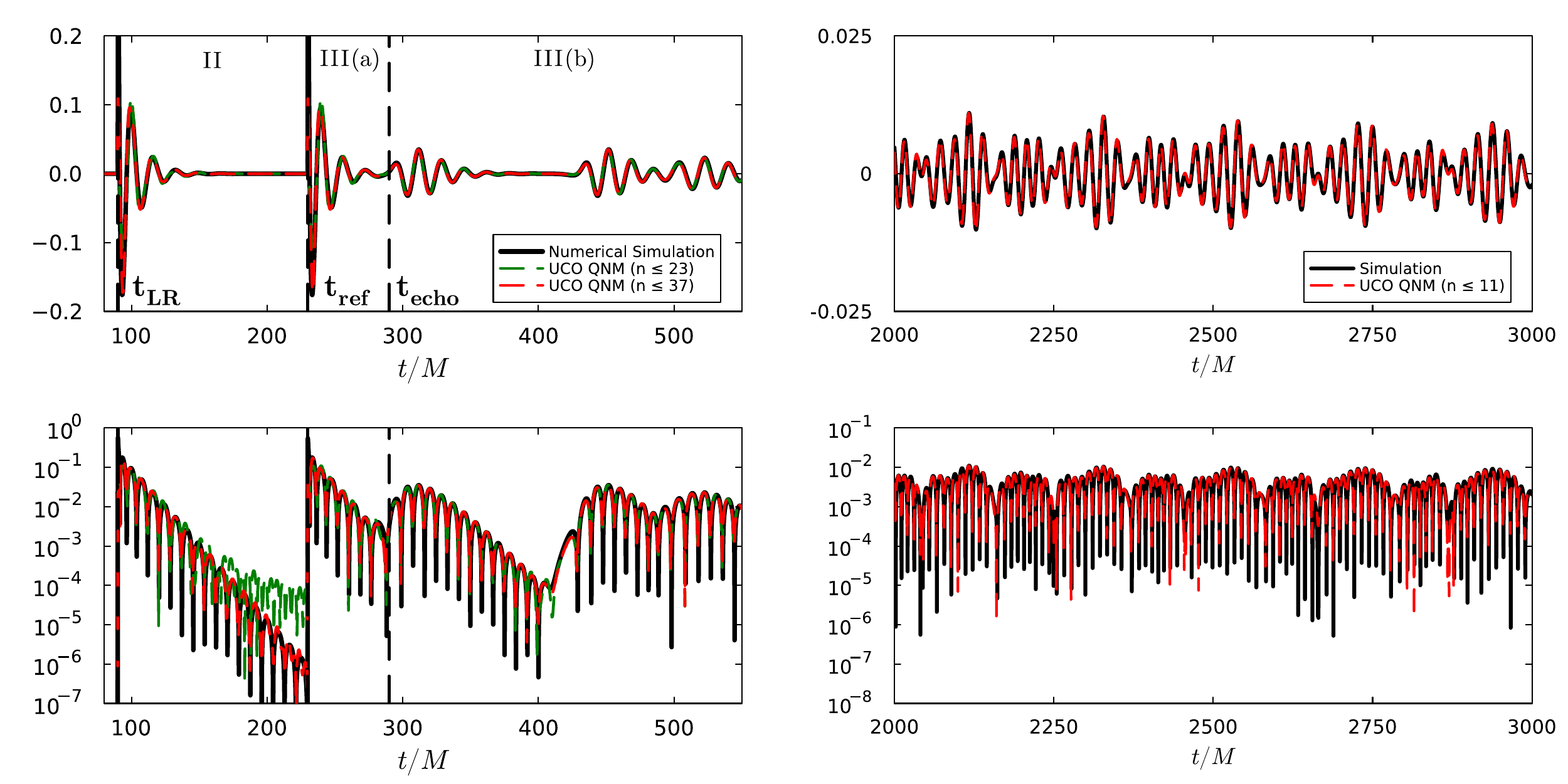}
    \caption{Detailed comparison of UCO Green function $G_\uco(x_\obs,x_s,t)$ for the inside case. Left: early time evolution from Stage II to Stage III(b). The red and green dashed lines correspond to the UCO QNM reconstruction with modes $n=0,...,37$ and $n=0,...,23$, respectively. Right: late time evolution at $t\gg t_\echo(x_\obs,x_s)$. The red dashed line represents the UCO QNM reconstruction using only the long-lived modes $n=0,...,11$. The top and bottom rows display the results on linear and logarithmic scales, respectively.
}
    \label{fig:UCOQNMinsiden12} 
\end{figure}

Figure~\ref{fig:UCOQNMinsiden12} provides a detailed comparison of the Green function at various times. For the early-time signal, including high-frequency modes again helps to reproduce the correct interference pattern of the pulses. Unlike the outside case, the UCO QNMs converge much more rapidly here in capturing the interference dips. 
For the echo signal (Stage III), the first 24 modes already suffice, compared with 40 modes needed for the outside case. More strikingly, with just 38 modes the BH ringdown phase, including the abrupt transition to the first echo, can be reconstructed from the very beginning of Stage II.
This explicitly demonstrates that destructive interference can suppress the amplitude by many orders of magnitude. 
For the late-time evolution, as the high-frequency modes become sufficiently damped, the low-frequency trapped modes alone suffice to provide an accurate reconstruction, just as in the outside case.

\section{Discussion and conclusion}
\label{sec:summary}

The central puzzle we address in this work is the apparent tension between the spectral instability of QNMs and the stability of the early prompt ringdown expected from causality, in the context of quantum BHs or horizonless ultracompact objects (UCOs). We adopt the truncated BH as a simple phenomenological model for these objects (below we refer to all of them simply as UCOs). By combining the proper decomposition of the UCO Green function with the choice of integration contours dictated by causality, a point emphasized in recent methodological developments for BHs, we provide a rigorous and unified picture of how the time-domain Green function is built from BH and UCO QNMs. Specifically, we consider two different scenarios: a source outside the light-ring barrier and a source inside it.

Although the detailed decompositions differ for the two cases, several common lessons emerge from the theoretical framework of Sec.~\ref{sec:theory} and are supported by the numerical validation of Sec.~\ref{sec:validaiton}. First, once the curved spacetime is probed, the UCO QNM reconstruction always provides a faithful description of the time-domain Green function, until the UCO tails begin to dominate at very late times. Second, by further decomposing the UCO Green function into a BH part plus echo corrections, we prove that the early time response is exactly the BH ringdown; the echo corrections contribute nothing when the contour is properly closed. Consequently, the UCO QNM expansions are formally equivalent to the BH QNM and tail expansions at early times, while their practical convergence depends on the source location.

The two cases differ in their practical consequences. For the outside case, the UCO Green function separates into a direct wave piece identical to the BH result and a piece sensitive to the light ring potential barrier. The latter encodes the UCO QNMs and tails and governs the evolution after \(t_\lr\) (Fig.~\ref{fig:contour_outside}), forming the basis of “quantum BH seismology”; the full numerical evolution is shown in Fig.~\ref{fig:FullCompOUT}. In principle the early ringdown can also be reconstructed from UCO QNMs, but this is highly inefficient because a very large number of high-frequency modes, with exponentially large excitation factors, are required to achieve the correct destructive interference. The BH QNM and tail expansion, by contrast, is very efficient in this stage, and the equivalence of the two bases is demonstrated in Fig.~\ref{fig:contour_outside2}. After the delay $t_d$ the mode amplitudes become comparable at all frequencies, making the UCO QNMs a good basis for the echoes.

The inside source case is markedly different. The Green function again splits into two components (Fig.~\ref{fig:contour_inside}): one analogous to the BH part, the other representing the interior reflection of the source-emitted wave and thus absent for a BH. This second component starts to contribute at an earlier time  $t_\re$, generating an additional echo train before the delay $t_d$ and demonstrating that ``quantum BH seismology'' can have a richer structure. Because the inside case is dominated by low-frequency content, the excitation factors are already comparable at all frequencies from the very beginning, and the UCO QNMs provide a good expansion basis for the full evolution (Fig.~\ref{fig:FullCompIN}), albeit less efficient than the BH basis early on. 
At late times, after the high-frequency modes have decayed, both cases are dominated by the long-lived trapped modes of the UCO.

These results answer the question raised in the Introduction about the meaning of ``BH spectroscopy'' and ``quantum BH seismology'' when near-horizon corrections exist. Standard BH spectroscopy verifies that the background spacetime is effectively that of a BH, but only up to the timescale at which interior reflections become relevant. Quantum BH seismology, on the other hand, is always applicable in principle, although the efficiency of the UCO QNM basis depends strongly on the time interval and the source location. At early times, a substantial destructive interference among high-frequency UCO modes is required to mimic the BH ringdown, whereas at late times a small subset of trapped modes suffices to capture the echo signature.

These findings complement and clarify the existing discussion in the literature. For the outside source case, Ref.~\cite{Rosato:2025lxb} showed that the UCO QNM reconstruction accurately recovers the echoes after the time delay $t_d$, while the early prompt ringdown was taken to be the BH ringdown without derivation from first principles. The formal equivalence of the two bases at early times was explicitly demonstrated in a different setting by Ref.~\cite{Oshita:2025ibu}, though it required a very large number of modes, and the underlying reason for the equivalence was not examined.  More recently, Ref.~\cite{OuldElHadj:2026zpp} explored a geometric series expansion of the echo part and analyzed in depth the relation between UCO QNMs and the Debye QNMs associated with each term of the series. The leading Debye contribution is exactly the BH QNMs and tail we studied here; focusing on the outside case, they encounter the same convergence difficulty with the UCO QNM reconstruction that we have described. They also show that higher-order Debye QNMs provide an alternative basis for echoes, albeit with an efficiency that gradually degrades with time. Importantly, the Debye QNM contributions are non-local in time, indicating that the time-domain Green function up to a finite time is not given by a finite truncation of the geometric series, contrary to what simple delta-function toy models suggest~\cite{Hui:2019aox}. 
This is consistent with the observation in Ref.~\cite{Daghigh:2025wcw}: the UCO QNMs disappear if only a finite number of terms in the series are kept, and retaining the full expression naturally restores them as the relevant basis.
Refs.~\cite{Wang:2019rcf,Oshita:2020dox} have considered the inside source case, treating each echo pulse in one-to-one correspondence with a series expansion term.
Our analysis thus supplies the missing link: the proper decomposition of the Green function and the evaluation of each component via distinct contours dictated by causality, unifying the different observations into a single consistent framework.

Several directions deserve further investigation. On the theory side, we have examined the decomposition of the UCO Green function for sources located far outside or far inside the light ring potential barrier; extending it to arbitrary positions remains important, since a general initial condition involves the Green function at multiple locations. A general Green function would allow the impact of initial conditions on quantum BH seismology to be systematically assessed (see Refs.~\cite{Conklin:2019fcs, Chavda:2024awq} for related work). 

On the observational side, the practical efficiency of different QNM bases is crucial for model-agnostic echo searches.  
Since the prompt ringdown is described exactly by the BH QNMs and tail, deviations from interior reflection should be sought only after a characteristic delay. A pipeline targeting long-lived UCO QNMs at late times has been developed and applied to real data~\cite{Ren:2021xbe,Wu:2023wfv,Wu:2025enn}. The targeted late-time echo signature is robust and insensitive to the source location, enabling efficient searches for strongly reflecting UCOs. When early echoes dominate (e.g., weak reflection), the LVK collaboration has adopted a quasiperiodic burst search~\cite{LIGOScientific:2020tif,LIGOScientific:2021sio,LIGOScientific:2026wpt}. 
However, if the signal exhibits a more complicated echo morphology, such as the interference of two echo trains, such a strategy may easily lose sensitivity.
Instead, a template based on higher-order Debye QNMs may be more efficient for reconstructing the early pulses in general. Combining both QNM sets would then provide a complementary strategy covering a wide range of quantum BH scenarios.

\vspace{0.1cm}
\section*{Acknowledgements} 
\vspace{-0.1cm}

We would like to thank Junquan Su for discussions on the BH Green function paper~\cite{Su:2026fvj}. J.R. is supported in part by the National Natural Science Foundation of China (Grant No. 12275276).

\appendix

\section{Properties of the homogeneous solutions}\label{app:homosolution}

For a Schwarzschild black hole, the two linearly independent solutions satisfying the required boundary conditions are
\begin{eqnarray}
X_\bh^+(\omega,x)&\to&
\left\{\begin{array}{ll}
D^\bh_\tra(\omega) e^{-i\omega x}+D^\bh_\re(\omega) e^{i \omega x},& x\to-\infty\\
%D^\bh_\out 
e^{i \omega x},& x\to\infty\,, \end{array}
\right.\label{eq:SNEampp} \\
X_\bh^-(\omega,x)&\to&
\left\{\begin{array}{ll}
%A^\bh_\tra 
e^{-i\omega x},& x\to-\infty\\
A^\bh_\in(\omega) e^{-i \omega x}+A^\bh_\out(\omega) e^{i \omega x},& x\to\infty\,, \end{array}
\right.\label{eq:SNEampm}
\end{eqnarray}
where $A(\omega)$ and $D(\omega)$ denote the asymptotic amplitudes. 
Regarding their analytic properties, $X_\bh^+$ can be expressed as a spectral series containing the confluent hypergeometric function and then possesses a branch-cut (BC) along the negative imaginary axis, while $X_\bh^-$ is analytic along the imaginary axis, owing to its Jaffe series representation~\cite{Leaver:1986gd}. 
For later discussion, it is convenient to define another solution $X^{\prime +}_\bh(\omega, x)$ that is purely ingoing at spatial infinity,, i.e. $X^{\prime +}_\bh(\omega, x)\to e^{-i\omega x}$ when $x\to \infty$. Owing to the reality of the potential, this solution satisfies $X^{\prime +}_\bh(\omega, x)=[X^+_\bh(\omega^*, x)]^*$, which implies that $X^{\prime +}_\bh$ carries a BC along the positive analytic axis.

%For conciseness, the explicit $\omega$ dependence of these amplitudes is suppressed in the notation.

For a spherically symmetric UCO, the solution that matches the outgoing condition at spatial infinity remains unchanged, while the ingoing solution is modified to
\begin{eqnarray}\label{eq:SNEamppUCO}
%X_{lm\omega}
X_\uco^-(\omega,x)\to
\left\{\begin{array}{ll}
e^{-i \omega x}+A^\uco_\re(\omega) e^{i \omega x},& x\to-\infty\\
A^\uco_\in(\omega) e^{-i \omega x}+A^\uco_\out(\omega) e^{i \omega x},& x\to\infty\,, \end{array}
\right.\,.
\end{eqnarray}
Assuming the interior reflection occurs near a radius with tortoise coordinate $x_0$, we parametrize the reflection at the interior boundary as
\begin{eqnarray}\label{eq:Aucoref}
    %\frac{A^\uco_\re}{A^\uco_\tra}
    A^\uco_\re=R_{\wa}(\omega)e^{-2i\omega x_0}\,,
\end{eqnarray}
where the dominant phase factor has been extracted, so that $R_{\wa}(\omega)$ contains only a small residual phase.

The amplitudes are interrelated. A convenient way to derive these relations is through the Wronskian of two linearly independent solutions $X(\omega, x)$ and $Y(\omega, x)$ 
\begin{eqnarray}
    W(X, Y)=Y\frac{\partial X}{\partial x}-X\frac{\partial Y}{\partial x}\,.
\end{eqnarray}
From the equation of motion,  $W(X, Y)$ depends only on $\omega$ and is independent of $x$. Evaluating the Wronskian at both asymptotic boundaries therefore yields simple algebraic relations among the various amplitudes.

Let's first consider the BH case. From the Wronskian $W(X^+_\bh, X^-_\bh) = 2i\omega A_\in^\bh(\omega) = 2i\omega D_\re^\bh(\omega)$, we obtain the relation $A_\in^\bh(\omega) = D_\re^\bh(\omega)$.  
The BH QNMs are therefore given by the zeros of $A_\in^\bh(\omega)$ (or, equivalently, $D_\re^\bh(\omega)$).  
From the analytic structure of $X^+_\bh$ and $X^-_\bh$, it follows that $A_\in^\bh(\omega)$ possesses a branch cut (BC) along the negative imaginary axis.   The transmission coefficient of the light-ring potential barrier can then be defined as 
\begin{eqnarray}\label{eq:TBH}
    \frac{1}{D^\bh_\re(\omega)}=\frac{1}{A^\bh_\in(\omega)}\equiv T_\bh(\omega)\,,
\end{eqnarray}
which inherits the same BC along the negative imaginary axis.
From $W(X^{\prime +}_\bh, X^-_\bh)$, we obtain the relation $A^\bh_\out(\omega)=-[D^\bh_\tra(\omega^*)]^*$.  %$D^\bh_\tra(\omega)=-[A^\bh_\out(\omega^*)]^*$. 
The analytic properties of $X^{\prime +}_\bh$ and $X^-_\bh$ imply that $A^\bh_\out(\omega)$ carries a BC along the positive imaginary axis, while  $D^\bh_\tra(\omega)$ possesses a BC along the negative imaginary axis. Since the reflection takes place near the peak of the potential barrier at $x_\lr$, it is convenient to factor out the $x_\lr$ dependence in the BH amplitudes, i.e.    
\begin{eqnarray}
D^\bh_\tra(\omega)=\bar{D}^\bh_\tra(\omega)e^{2i\omega x_\lr},\quad A^\bh_\out(\omega)=\bar{A}^\bh_\out(\omega)e^{-2i\omega x_\lr}\,.
\end{eqnarray}
The reflection coefficient of the potential barrier is then defined as
\begin{eqnarray}\label{eq:RBH}
    \frac{\bar{D}^\bh_\tra(\omega)}{D^\bh_\re(\omega)}
    =-\frac{[\bar{A}^\bh_\out(\omega^*)]^*}{A^\bh_\in(\omega)}
    %=-\frac{(A^\bh_\out/A^\bh_\tra)^*}{(A^\bh_\in/A^\bh_\tra)}
    \equiv R_{\bh}(\omega)\,. %e^{2i\omega x_{\rm LR}}\,.
\end{eqnarray}
which no longer contains the artificially large $x_\lr$-dependent phase. This quantity also possesses a BC along the negative imaginary axis.
From $W(X^+_\bh, X^{\prime +}_\bh)$, we obtain the normalization conditions,
\begin{eqnarray}
D^\bh_\re(\omega)[D^\bh_\re(\omega^*)]^*-D^\bh_\tra(\omega)[D^\bh_\tra(\omega^*)]^*=1\,.
\end{eqnarray}
When $\omega$ lies on the real axis, it reduces to the standard unitarity relation for the transmission and reflection coefficients: $|R_\bh(\omega)|^2+|T_\bh(\omega)|^2=1$.

Now we relate the UCO amplitudes to the BH ones. From $W(X^+_\bh, X^-_\uco)$, we find 
\begin{eqnarray}\label{eq:Rel10}
    A^\uco_\in(\omega)=D^\bh_\re(\omega) - D^\bh_\tra(\omega) A^\uco_\re(\omega)\,.
\end{eqnarray}
If $\omega$ sits at any QNMs of UCO, i.e. $A^\uco_\in(\omega_n)=0$, Eq.~(\ref{eq:Rel10}) reduces to $D^\bh_\re(\omega_n) A^\uco_\tra(\omega_n)=D^\bh_\tra(\omega_n) A^\uco_\re(\omega_n)$, i.e.
\begin{eqnarray}\label{eq:UCOQNMs_app}
   R_{\bh}(\omega_n)R_\wa(\omega_n) e^{i\omega_n t_d}=1\,,
\end{eqnarray}
which is precisely the condition for UCO QNMs in Eq.~(\ref{eq:UCOQNMs}). 
Away from the QNMs of both the BH and UCO, i.e. $A^\bh_\in(\omega)\neq0$ and $A^\uco_\in(\omega)\neq0$, this gives
\begin{eqnarray}\label{eq:Rel11}
    \frac{1}{A^\uco_\in(\omega)}=\frac{1}
    %{D^\bh_\out}
    {D^\bh_\re(\omega)}\left(1-\frac{D^\bh_\tra(\omega)}{D^\bh_\re(\omega)}A^\uco_\re(\omega)\right)^{-1}
    =\frac{1}{A^\bh_\in(\omega)}\Big(1-R_{\bh}(\omega)R_\wa(\omega) e^{i\omega t_d}\Big)^{-1}\,,
\end{eqnarray}
where $t_d\equiv 2(x_{\rm LR}-x_0)$ is the round-trip light-travel time between the light-ring peak and the interior boundary. 
%If $\omega$ sits at any QNMs of BH, i.e. $A^\bh_\in=0$, the same equation gives  $D^\bh_\out A^\uco_\in=- D^\bh_\tra A^\uco_\re$ (??). 
From $W(X^-_\bh, X^-_\uco)$, we obtain 
\begin{eqnarray}\label{eq:Rel20}
    A^\bh_\out(\omega) A^\uco_\in(\omega)-A^\bh_\in(\omega) A^\uco_\out(\omega)
    =- A^\uco_\re(\omega)\,.
\end{eqnarray}
At the UCO QNMs, it yields 
\begin{eqnarray}
    A^\uco_\out(\omega_n)= \frac{A^\uco_\re(\omega_n)}{A^\bh_\in(\omega_n)}\,.
\end{eqnarray}
Away from any QNMs, we can find
\begin{eqnarray}\label{eq:Rel21}
    \frac{A^\uco_\out(\omega)}{A^\uco_\in(\omega)}=\frac{A^\bh_\out(\omega)}{A^\bh_\in(\omega)}+\frac{1}{A^\bh_\in(\omega)}\frac{A^\uco_\re(\omega)}{A^\uco_\in(\omega)}
    =\frac{A^\bh_\out(\omega)}{A^\bh_\in(\omega)}
    +\frac{T_\bh(\omega) R_\wa(\omega) e^{-2i\omega x_0}}{A_\in^\uco(\omega)}\,.
\end{eqnarray}
Assuming an analytic interior reflection coefficient $R_\wa(\omega)$, the difference between the UCO and BH amplitude ratios possesses a BC only along the negative imaginary axis.
Finally, we define another solution $X^{\prime -}_\uco(\omega,x)=[X^{-}_\uco(\omega^*,x)]^*$, which also satisfies the original field equation. From $W(X^-_\uco, X^{\prime -}_\uco)$, we obtain the normalization condition for the UCO amplitudes,
\begin{eqnarray}
A^\uco_\out(\omega)[A^\uco_\out(\omega^*)]^*-A^\uco_\in(\omega)[A^\uco_\in(\omega^*)]^*=A^\uco_\re(\omega)[A^\uco_\re(\omega^*)]^*-1\,.
\end{eqnarray}
When $\omega$ is real, this reduces to the ``energy conservation'' relation: $|A^\uco_\out(\omega)|^2-|A^\uco_\in(\omega)|^2=|R_\wa(\omega)|^2-1$.

\section{Initial condition dependence}\label{app:initialcond}

The time-domain Green function corresponds to an idealized $\delta$-function source (or initial condition). In our numerical simulations we approximate it by a narrow Gaussian profile. In this appendix we examine how the chosen initial profile affects that approximation, with particular attention to its impact on the QNM reconstruction.   
Specifically, we consider initial data with a vanishing field value and a narrow Gaussian time derivative. From Eq.~(\ref{eq:timedomwf}) the resulting time-domain waveform at $x=x_\obs$ is then
\begin{equation}\label{eq:psitx2}
    \psi(t,x_\obs)
    =
    -\int_{-\infty}^{\infty} d x'\,
    G(t,x_\obs,x')\,\partial_t\psi(0,x')\,,
\end{equation}
where we set $t_0=0$. The initial velocity profile is: $\partial_t\psi(0,x') = \frac{1}{\sqrt{2\pi}\sigma} \exp[-\frac{1}{2\sigma^2}(x'-x_s)^2]$, with a sufficiently narrow width $\sigma=0.01M$.

The QNM contribution to the waveform can be extracted from the QNM part of the Green function. Without loss of generality, the QNM contribution to the Green function can be written as
$G_\uco^{\rm QNM}(t,x,x')=
\sum_n B_n e^{-i\omega_n(t-t_I(x,x'))}$, where $B_n$ are the excitation factors discussed earlier, $t_I(x,x')$ is the characteristic time scale used to define $B_n$, and the sum runs over both positive- and negative-frequency modes. Substituting this decomposition into Eq.~(\ref{eq:psitx2}) yields 
\begin{eqnarray}
    \psi_{\rm QNM}(t,x_\obs)
    &\approx &
    -\int_{x_s-\delta_-(t)}^{x_s+\delta_+(t)}
    dx'\,
    \sum_n B_n e^{-i\omega_n(t-t_I(x_\obs,x'))}
    \frac{1}{\sqrt{2\pi}\sigma}
    \exp\!\left[-\frac{(x'-x_s)^2}{2\sigma^2}\right]\nonumber\\
    &=& -\sum_n B_n e^{-i\omega_n(t-t_I(x_\obs,x_s))}\frac{1}{\sqrt{2\pi}\sigma}\int^{\delta_+(t)}_{-\delta_-(t)} dy \, e^{\pm i\omega_n y}e^{-\frac{1}{2\sigma^2}y^2}\nonumber\\
    &=& \sum_n \mathcal{C}^\pm_n(t) e^{-i\omega_n(t-t_I(x_\obs,x_s))}\,,
\end{eqnarray}
where $y\equiv x'-x_s$. The quantities $\mathcal{C}^\pm_n(t)$ denote the QNM excitation coefficients corresponding to the two possible signs of the $x'$-dependence in the phase factor, i.e. $\pm i \omega_n x'$. $\delta_+(t)$ and $\delta_-(t)$ determine the upper and lower boundaries of the $x'$-integration range over which the approximation applies. Because the QNM representation of $G_\uco$ is only valid within finite time intervals, these integration limits can be time-dependent. Consequently, the QNM excitation coefficients $\mathcal{C}^\pm_n(t)$ can also become time-dependent~\cite{Chavda:2024awq}, and are given by 
\begin{eqnarray}
    \mathcal{C}^\pm_n(t)=B_n\, e^{-\frac{1}{2}(\sigma \omega_n)^2}\mathcal{E}_n(t),\;\;
    \mathcal{E}_n(t)=\frac{1}{2}\left[\erf\left(\frac{\delta_+(t)}{\sqrt{2}\sigma}\mp i \frac{\sigma \omega_n}{\sqrt{2}}\right)+\erf\left(\frac{\delta_-(t)}{\sqrt{2}\sigma}\pm i \frac{\sigma \omega_n}{\sqrt{2}}\right)\right]\,,
\end{eqnarray}
where $\erf(z)$ denotes the error function arising from the finite integration range. 
For $\mathcal{C}_n^\pm(t)$ to faithfully approximate the original $B_n$, i.e. $\mathcal{E}_n(t)\approx 1$, the following conditions must be satisfied: 
\begin{eqnarray}\label{eq:sigmacond}
    |\omega_{n,R}|,|\omega_{n,I}|\ll 1/\sigma,\quad 
    \delta_\pm(t)\gg \sigma\,.
\end{eqnarray}
The first condition imposes an upper bound on the QNM frequencies that can be included in the sum; beyond this bound, the QNM representation of the Green function no longer matches the physical waveform. For our numerical study, we restrict to modes with $M|\omega_{n,R}|, M|\omega_{n,I}|\ll 10$, so that with $\sigma=0.01M$ the requirements $|\omega_{n,R}|\sigma, |\omega_{n,I}|\sigma\ll 1$ are always satisfied.

The second condition restricts the approximation to a finite time window. To illustrate it more concretely, we examine the QNM reconstruction cases discussed in the main text. 
For the outside source case, consider the QNM representation in Stage III in Fig.~\ref{fig:contour_outside}, i.e. the first equality in Eq.~(\ref{eq:GFtimeOUT1}). It applies only when $t>t_\lr(x_\obs,x')$ in Eq.~(\ref{eq:tlr}); this imposes an upper bound on the $x'$-integration range, yielding
\begin{eqnarray}
    \delta_+(t)
    %=t-(\bar{x}+\bar{x}_s)
    =t-t_\lr(x_\obs,x_s)\,.
\end{eqnarray}
The lower bound follows from the requirement $x'\gg x_\lr$, and leads to the time-independent condition $\delta_-\ll \bar{x}_s$. The condition in Eq.~(\ref{eq:sigmacond}) then implies that the approximation is accurate only when $t-t_{\lr}(x_\obs,x_s)\gg \mathcal{O}(1)\sigma$. For our benchmark, with $t_\lr(x_\obs,x_s)\sim \mathcal{O}(10)M$ and $\sigma=0.01M$, the constant-coefficient QNM reconstruction becomes valid almost immediately after $t_\lr(x_\obs,x_s)$, and would not affect the results presented in the figures. Moreover, because $\bar{x}_s\gg \sigma$, t is always possible to choose a lower bound such that $\sigma\ll\delta_-\ll\bar{x}_s$ and the corresponding error function approaches unity.

For the inside source case, the QNM Green function contains two causal components, as shown in Fig.~\ref{fig:contour_inside}. The first component, corresponding to the excitation factor $\bar{B}_n^+$ in Eq.~(\ref{eq:GFtimeIN1}), becomes relevant when $t>t_\lr(x_\obs,x')$  in Eq.~(\ref{eq:tlrIN}), which gives the lower bound for the $x'$-integration,
\begin{eqnarray}
    \delta_-(t)
    %=t-(x-x_s)
    =t-t_\lr(x_\obs,x_s)\,.
\end{eqnarray}
The upper bound follows from  $x'\ll x_\lr$, and is time-independent, $\delta_+=-\bar{x}_s$. For the second component associated with $\tilde{B}_n^{++}$ in Eq.~(\ref{eq:GFtimeIN2}), the causality condition $t>t_\re(x_\obs,x')$ in Eq.~(\ref{eq:tref}) together with $x'\ll x_\lr$ yields the time-dependent upper bound,
\begin{eqnarray}
    \delta_+(t)
    %=\min\{t-(x+x_s-2x_0),-\bar{x}_s\}
    =\min\{t-t_\re(x_\obs,x_s),-\bar{x}_s\}\,,
\end{eqnarray}
while there is no restriction on the lower boundary. Therefore, from Eq.~(\ref{eq:sigmacond}), we have $\mathcal{C}^\pm_n\approx B_n$ when $t-t_\lr(x_\obs,x_s), t-t_\re(x_\obs,x_s)\gg \mathcal{O}(1)\sigma$ and $\sigma\ll\delta_+\ll-\bar{x}_s$. 

In summary, the initial Gaussian profile is chosen sufficiently narrow compared to the source locations and the frequency modes retained in the QNM reconstruction, in both the outside and inside cases. As a result, the constant-coefficient QNM reconstruction with the original excitation factors matches the time-domain waveform at all times, except for narrow intervals of width $\sim\mathcal{O}(\sigma)$ around the characteristic times $t_\lr(x_\obs,x_s)$ and $t_\re(x_\obs,x_s)$.

\clearpage
\newpage

\bibliographystyle{unsrt}
\bibliography{ref}

%\bibliographystyle{unsrt}
%\nocite{*}
%\bibliography{reference}

\end{document}